\documentclass[reprint,aip,jcp,superscriptaddress,floatfix]{revtex4-1}

\usepackage{amsmath,amssymb}
\usepackage{graphicx}
\usepackage{color}
\usepackage{soul}
\usepackage{ulem}
\usepackage{lineno}

\allowdisplaybreaks

\renewcommand{\vec}[1]{\boldsymbol{#1}}
\renewcommand{\Re}{\operatorname{Re}}

\newcommand{\avg}[1]{\left\langle #1 \right\rangle}
\newcommand{\kT}{k_\text{B}T}
\newcommand{\omb}{\omega_\text{b}}

\newcommand{\TST}{\textnormal{TST}}

\newsavebox\CBox
\newcommand\hcancel[2][0.5pt]{{\color{red}{%
  \ifmmode\sbox\CBox{$#2$}\else\sbox\CBox{#2}\fi%
  \makebox[0pt][l]{\usebox\CBox}%
  \rule[0.5\ht\CBox-#1/2]{\wd\CBox}{#1}}}}

\graphicspath{{figs/}}

\begin{document}


\title{Rate calculation with correlated noise}
\author{Thomas Bartsch}
\affiliation{Department of Mathematical Sciences, Loughborough University,
Loughborough LE11 3TU, United Kingdom.}
\author{F. Revuelta}
\affiliation{Grupo de Sistemas Complejos,
Escuela T\'ecnica Superior de Ingenier\'ia Agron\'omica, Alimentaria
y de Biosistemas,
Universidad Polit\'ecnica de Madrid,
Avda.\ Complutense s/n, 28040 Madrid, Spain.}
\affiliation{Instituto de Ciencias Matem\'aticas (ICMAT), 
Cantoblanco, 28049  Madrid, Spain.}
\author{R. M. Benito}
\affiliation{Grupo de Sistemas Complejos,
Escuela T\'ecnica Superior de Ingenier\'ia Agron\'omica, Alimentaria
y de Biosistemas,
Universidad Polit\'ecnica de Madrid,
Avda.\ Complutense s/n, 28040 Madrid, Spain.}
\author{F. Borondo}
\affiliation{Instituto de Ciencias Matem\'aticas (ICMAT), 
Cantoblanco, 28049  Madrid, Spain.}
\affiliation{Departamento de Qu\'imica, 
Universidad Aut\'onoma de Madrid, Cantoblanco, 28049  Madrid, Spain.}

\date{\today}

\begin{abstract}
The usual identification of reactive trajectories for the calculation 
of reaction rates requires very time-consuming simulations, 
particularly if the environment presents memory effects. 
In this paper, we develop a new method that permits the identification 
of reactive trajectories in a system under the action of a stochastic 
colored driving. 
This method is based on the perturbative computation of the invariant 
structures that act as separatrices for reactivity. 
Furthermore, using this perturbative scheme, 
we have obtained a formally exact expression for the reaction rate 
in multidimensional systems coupled to colored noisy environments.
\end{abstract}

\pacs{82.20.Db, 05.40.Ca, 05.45.2a, 34.10.+x}

\maketitle

\section{Introduction}
\label{sec:Intro}

Ever since its inception \cite{Marcelin15,Evans35,Wigner38,Horiuti38} 
Transition State Theory (TST) has provided a powerful conceptual framework 
for reaction rate theory.
Though originally devised to describe chemical reactions of small molecules, 
it has been
applied to a wide variety of activated processes, that proceed 
from suitably defined ``reactant'' to ``product'' states%
\cite{Toller85,Eckhardt95,Hernandez93,Hernandez94,Jaffe99,Jaffe00,Koon00,%
Jaffe02,Uzer02,Komatsuzaki02}.
In all these cases, TST identifies the rate limiting step of the reaction, 
and thus the reaction mechanism, and provides a simple approximation to 
the reaction rate.

More precisely, TST applies to systems in which the rate limiting step 
is the crossing of an energetic barrier.
In this situation the vast majority of reactive trajectories will pass 
very close to the top of the barrier. 
If a dividing surface (DS) between reactant and product regions of 
phase space is chosen close to the barrier top, the reaction rate 
can be computed from the steady-state flux of trajectories through 
this surface.
To avoid overestimating the rate, one must ensure that trajectories 
are only included in the flux calculation if they are actually reactive.
The identification of reactive trajectories requires, in principle,
a study of the reaction dynamics in all its complexity.
A simple method to perform this central task is therefore highly desirable.

The crudest
approach to identify reactive trajectories is to assume 
that every trajectory that crosses the DS from the reactant to the 
product side is reactive.
This approximation, which leads to the standard TST~rate formula, 
is equivalent to the postulate that no trajectory can cross the 
DS more than once.
Depending on the choice of DS, the TST~approximation can be more or 
less accurate.
However, for gas phase reactions and energies close to the reaction threshold, 
a DS can be constructed that is rigorously recrossing free and therefore 
TST renders exact rates \cite{Uzer02,Waalkens04a,Waalkens04c}.
A recrossing free DS cannot be found at 
higher energies 
\cite{Pechukas79,Pollak80,Allahem12} or if the reactive system is 
strongly coupled to an environment, for example a liquid solvent.
In the latter case, in particular, a typical trajectory will cross and recross 
any given DS many times, so that TST will grossly overestimate the reaction rate.
For this reason, much effort has been invested into the construction of a 
DS that minimizes recrossings (see Ref.~\onlinecite{Garrett05a} for a review).

Since for reactions in solution the recrossing problem cannot be overcome 
by a suitable choice of DS, other methods must be sought.
A numerical simulation of trajectories does, of course, 
provide a reliable criterion, but it can be computationally very demanding.
The simplicity of TST suggests that a more straightforward criterion might be available.
For reactive systems coupled to a heat bath modeled by a Langevin 
equation (LE), we recently suggested such a criterion~\cite{Revuelta12,Bartsch12}:
the trick is to shift focus away from the DS onto
hypersurfaces in phase space that separate reactive from non-reactive 
trajectories.
These surfaces, which are invariant manifolds, can be characterized 
dynamically from a more fundamental point of view,
and this characterization allows one to compute them.
In the present work we generalize the method of Refs.~\onlinecite{Revuelta12,Bartsch12}
to a heat bath modeled by a generalized Langevin equation (GLE),
which takes the finite relaxation time of the bath into account.
Some of these results have already been reported in Ref.~\onlinecite{Revuelta16}.
We will here supply the missing details
and extend the result to multidimensional systems.

The usual LE has been widely used to model the interaction of a 
reactive system with a surrounding heat bath.\cite{Haenggi90,Pechukas76,Chandler78}
It neglects quantum effects such as barrier tunneling, which can be important 
in the case of light particles \cite{Bothma10}, and the interaction with 
electronic excited states through conical intersections \cite{Polli10}.
More importantly for our purposes, it also neglects the internal dynamics of 
the heat bath and assumes instead that the bath equilibrates infinitely fast.
A more realistic model of a heat bath will take into account that the 
heat bath molecules need a finite time to move.
As a consequence, the stochastic forces that the bath exerts on the reactive 
system at different times must be correlated, and this correlation will decay 
on a time scale that is given by the dynamics of the bath.
This effect can be described by a GLE
(see Refs.~\onlinecite{Haenggi90,Pechukas76,Chandler78} 
and Sect.~\ref{sec:GLE} below).

The original LE as well as its generalization for correlated 
noise are equivalent to a Hamiltonian model in which the reactive system is 
coupled to a bath of infinitely  many harmonic oscillators.\cite{Zwanzig73}
Via this representation, the rate theory originally developed 
by Kramers\cite{Kramers40} for white (uncorrelated) noise 
and by Grote and Hynes\cite{Grote80} for colored (correlated) noise, 
can be obtained from a TST in an infinite-dimensional phase space.\cite{Pollak86}
This approach could then be extended to include the corrections due 
to anharmonic barriers.\cite{Pollak93a,Talkner93,Talkner94a}
In this work we  avoid using an explicit model of the heat bath that 
introduces an infinite-dimensional phase space.
Instead, we  work directly in the phase space of the GLE,
which is finite-dimensional for the friction kernels we consider.
This choice is convenient both from a computational
and from a conceptual points of view,
since it allows to visualize the relevant phase space structures more easily.

In this paper, we present a detailed study of the phase space structures 
of the GLE introduced in Ref.~\onlinecite{Revuelta16}
that determine reactivity.
This work is based on a recent series of papers 
\cite{Bartsch05b,Bartsch05c,Bartsch06a,Bartsch08,Hernandez10,Revuelta12,
Bartsch12,Revuelta16} 
that describe such structures and their use in rate theory, 
including the identification of reactive trajectories\cite{Bartsch06a} 
and rate calculation\cite{Bartsch08,Revuelta12,Bartsch12, Revuelta16}.
With the exception of Refs.~\onlinecite{Revuelta16} and~\onlinecite{Bartsch05c}, 
the previous papers were restricted to the LE with white noise.
They show that the LE gives rise to a particular trajectory called the 
Transition State (TS) trajectory that remains in the vicinity of the 
barrier top for all times, 
without ever descending into either well.
It depends on the realization of the noise and takes over the role played 
by the saddle point in the TST of autonomous systems.
For the case of a harmonic barrier, 
it was shown in Refs.~\onlinecite{Bartsch05b,Bartsch05c} 
that the LE becomes noiseless if the dynamics is studied 
in a time-dependent coordinate system with the TS~trajectory as the origin.
It is then easy to identify a recrossing free DS in the moving coordinate system, 
as well as hypersurfaces that separate reactive from nonreactive trajectories.
The most important of these surfaces is the stable manifold of the TS~trajectory.
It contains all trajectories that asymptotically approach the TS~trajectory 
for long times.
This stable manifold separates trajectories that descend into the product 
well in the distant future from those that descent into the reactant well.
A knowledge of the stable manifold therefore allows one to distinguish reactive 
from nonreactive trajectories without any further computation.
It solves the diagnostic problem that is fundamental to rate theory.

The stable manifold will persist if the barrier is not harmonic. 
In Refs.~\onlinecite{Revuelta12,Bartsch12} we demonstrated how it can be 
computed by perturbation theory.
As already announced in  Ref.~\onlinecite{Revuelta16}, 
we will demonstrate here in detail that the stable manifold 
also exists in a reactive system  described by a 
GLE with correlated noise, and we will use it to derive anharmonic 
barrier corrections to the reaction rate for such systems.
In the first part of the paper, 
in Secs.~\ref{sec:GLE}-\ref{sec:kappa},
we consider one-dimensional systems.
Leading order rate corrections are derived for a generic one-dimensional 
barrier potential.
For an application to a  realistic system, 
i.e.~LiNC$\rightleftharpoons$LiCN isomerization, see
Ref.~\onlinecite{Revuelta16}.
In the second half of the paper, in Sec.~\ref{sec:2d},
we extend  the same computational method 
to multidimensional systems.
We then derive the first and second order rate corrections for the 
anharmonic two-dimensional model potential that was already used in 
Refs.~\onlinecite{Bartsch06a,Bartsch08,Revuelta12,Bartsch12}.

The outline of the paper is as follows. 
In Sect.~\ref{sec:rate} we introduce the fundamentals of the rate theory 
that are necessary for our purposes. 
Section~\ref{sec:GLE} presents the GLE and its phase space coordinates. 
The geometrical structures that characterize the phase space of a system 
of one degree--of--freedom (dof) and are central for our study are 
described in Sect.~\ref{sec:mf}. 
Section~\ref{sec:vCrit} is devoted to the calculation of a critical velocity 
that allows a unique identification of reactive trajectories. 
In Sect.~\ref{sec:kappa} we explain how this critical velocity can be used 
for the calculation of the transmission factor. 
Finally, we summarize in Sect.~\ref{sec:conclu} the conclusions of our work.

\section{The fundamental rate formula}
\label{sec:rate}
In this section, we summarize the fundamentals of reaction rate theory 
that will be used in the rest of the paper. 
For a more detailed discussion, 
see for example Refs.~\onlinecite{Haenggi90, Pechukas76, Chandler78}.

As mentioned above, TST is based on the assumption that there is 
a recrossing free DS between reactants and products, 
that is crossed once and only once by every reactive trajectory.
If we assume that this DS is placed at~$x^\ddag$ and that 
the reactant and product regions are defined by~$x<x^\ddag$ and~$x>x^\ddag$,
respectively, the TST approximation to the reaction rate is given  by the 
flux-over-population expression
\begin{equation}
   \label{kexact}
   k=\frac{J}{N},
\end{equation}
where~$N$ is the average population of the reactant region 
and~$J$ is the reactive flux out of it.
In a system with $n$ dof, the DS $x=x^\ddag$ can be parameterized 
by $2n-1$ phase space coordinates: 
the velocity $v_x$ perpendicular to the surface and the coordinates 
$\vec q_\perp$ and corresponding velocities $\vec v_\perp$ 
in the transverse directions.
The reactive flux is then given by
\begin{equation}
   \label{Jexact}
   J=\avg{v_x \, \chi_\alpha(v_x,\vec q_\perp, \vec v_\perp)}_{\alpha,\textnormal{IC}},
\end{equation}
where the average extends over all realizations $\alpha$ of the noise and over a
stationary-state ensemble of initial conditions (IC's) on the~DS.
The characteristic function $\chi_\alpha$ takes now a value equal to~1 
if the trajectory given by the IC~$(x^\ddag, v_x, \vec{q}_\perp, \vec{v}_\perp)$ 
is reactive if driven by the noise sequence $\alpha$ and~0 otherwise.
It ensures that a trajectory is only included in the reactive flux 
if it actually leads to a reaction, i.e., if it descends from the barrier 
into the product region and thermalizes there.
The main dynamical challenge in a rate calculation consists in the evaluation 
of the characteristic function $\chi_\alpha$.
We will later propose a simple explicit expression for $\chi_\alpha$, 
in Eq.~\eqref{chi_r}, that concentrates the potentially intricate dynamics 
of the system into a single function.

Standard TST sidesteps the dynamical problem by assuming 
that the DS is recrossing free.
It then follows that a trajectory that crosses the DS with a positive velocity~$v_x$ 
will move from the reactant to the product side, contributing to the reactive flux,
whereas a trajectory with a negative~$v_x$ will end in the reactant side and, 
as a consequence, will be nonreactive.
In other words, TST assumes the characteristic function 
\begin{equation} \label{chi}
  \chi^\TST(v_x)=\left\{
                  \begin{array}{cc}
                    1, & \textnormal{if }  v_x>0, \\
                    0, & \textnormal{if }  v_x<0 .\\
                  \end{array}
                 \right.
\end{equation}
In the corresponding flux
\begin{equation}
  \label{Jtst}
   J^\TST=\avg{v_x \, \chi^\TST(v_x)}_{v_x},
\end{equation}
the average only needs to be extended over the velocity $v_x$ because the 
argument is independent of all other coordinates and of the noise.

The standard TST~approximation to the reaction rate 
\begin{equation}
   \label{ktst}
   k^\text{TST}=\frac{J^\text{TST}}{N},
\end{equation}
always overestimates the true rate.
The extent to which a given system violates the no-recrossing 
assumption is measured by the transmission factor 
\begin{equation}
   \label{kappa}
   \kappa=\frac{k}{k^\TST}<1.
\end{equation}

Unless the friction caused by the heat bath is very weak, the stationary-state 
distribution of IC's in the barrier is given by a Boltzmann equilibrium distribution.
This assumption  will always be made in the rate calculations presented here,
though the dynamical theory at the heart of this study does not require it.
The exact expression~\eqref{chi_r} for the characteristic function applies 
to equilibrium as well as nonequilibrium systems.
The average over IC's is then performed over an ensemble with
probability density
\begin{equation}
    \label{prob}
    p(x, v_x, \vec{q}_\perp,
    \vec{v}_\perp)=\delta(x-x^\ddag) \exp\left( -\frac{m v_x^2}{2\kT} \right) 
                 p_\perp(\vec{q}_\perp, \vec{v}_\perp),
\end{equation}
where~$m$ is the particle mass and $p_\perp$ is a Boltzmann distribution
\begin{equation}
    \label{Pperp}
    p_\perp(\vec{q}_\perp, \vec{v}_\perp)=\frac{1}{Z}
          \exp\left( -\frac{m v_x^2/2+U(x^\ddag, \vec{q}_\perp)}{\kT} \right)
\end{equation}
for the transverse coordinates and velocities,
being~$U(x^\ddag, \vec{q}_\perp)$ the potential of mean force.
The factor~$Z$ in Eq.~\eqref{Pperp} is the partition function that ensures
\begin{equation}
    \int d\vec{q}_\perp d\vec{v}_\perp \, p_\perp(\vec{q}_\perp, \vec{v}_\perp) = 1.
\end{equation}
Under this assumption, the TST~flux~\eqref{Jtst} can be evaluated analytically 
to give
\begin{equation}
	J^\textrm{TST} = \sqrt{\frac{\kT}{2\pi\,m}}.
\end{equation}
The exact flux~\eqref{Jexact} is evaluated by randomly sampling 
IC's from the ensemble~\eqref{prob} and noise sequences.

To provide a benchmark for the perturbative calculations, classical trajectories
are numerically propagated, using
the algorithm described in Refs.~\onlinecite{Hershkovitz98, Hershkovitz01} 
until their energy is far enough below the saddle point,
so that they can be considered thermalized. 
As will be demonstrated below, the selection of this cutoff energy is 
much more critical for the correct computation of reaction rates 
in presence of colored noise than for the case of environments 
characterized by white noise.
Actually, no matter how low the value of the cutoff is chosen, 
some trajectories will always recross the DS if one waits long enough.
However, if the particle remains for long enough in the well into which it has descended, 
any further recrossing can be regarded as part of subsequent reaction events.

\section{The generalized Langevin equation}
\label{sec:GLE}

The reduced dynamics of an $n$--dof system coupled to an external heat bath 
that has memory effects can be accurately described by the GLE
\begin{equation}
	\label{genLE_I}
	m \ddot{\vec{q}} = -\nabla_{\vec{q}} U(\vec{q}) 
   - m \int_{-\infty}^t \mathbf{\Gamma}(t-s)\, \dot{\vec q}(s)\,ds + m \,\vec{R}_{\alpha}(t),
\end{equation}
where $m$ is the particle mass,
$\vec{q}$ is an $n$--dimensional coordinates vector, 
$\mathbf{\Gamma}(t)$ is the  friction kernel $n\times n$ matrix, 
and $\vec{R}_{\alpha}(t)$ is the fluctuating noise force exerted by the heat bath. 
Moreover, $\mathbf{\Gamma}(t)$ and~$\vec{R}_\alpha(t)$ are related to 
each other according to the fluctuation-dissipation theorem
\begin{equation}
        \label{eq:R0Rt}
	\avg{\vec{R}_\alpha(0) \vec{R}^{\rm T}_\alpha(t)}_\alpha = 
       \frac{\kT\,\mathbf{\Gamma}(t)}{m},
\end{equation}
where $\avg{...}_\alpha$ denotes an average over the different realizations~$\alpha$ 
of the noise.
In the first part of this paper, we will focus on the study of one-dimensional problems. 
In this case, the coordinate vector~$\vec{q}$ has a single component $x$, 
and Eq.~\eqref{genLE_I} reduces to
\begin{equation}
	\label{genLE}
	m \ddot x = - \frac{d U(x)}{d x} - m \int_{-\infty}^t \gamma(t-s)\, \dot x(s)\,ds 
                + m R_\alpha(t).
\end{equation}
The potential energy can be expanded as a Taylor series around its saddle point as
\begin{equation}
        \label{Uexpand}
	U(x) = -\frac{m\omb^2}{2}x^2 + \varepsilon \frac{m c_3}{3} x^3
           +\varepsilon^2 \frac{m c_4}{4} x^4+\ldots,
\end{equation}
where the formal perturbation parameter~$\varepsilon$ measures the strength of the 
anharmonicity that comes into play as the particle moves away from the saddle point.
It is only used to keep track of the expansion order 
and will be set equal to~1 at the end of the calculations.
Using this expansion, the mean force turns into
\[
	-\frac{d U(x)}{d x} = m\omb^2x + m\,f(x)
\]
with $f(x)=-\varepsilon c_3 x^2 - \varepsilon^2 c_4 x^3 - \dots$ denoting 
the anharmonic terms.

\subsection{The extended phase space}
\label{subsec:PS1D}

In this work, we assume an exponential friction kernel
\begin{equation}
	\label{gammaExp}
	\gamma(t) = \frac{\gamma_0}{\tau}\, e^{-t/\tau},
\end{equation}
with a characteristic correlation time~$\tau$ and a damping strength~$\gamma_0$.
It accurately describes the behavior of many realistic chemical reactions~\cite{Muller10}. 
In this case, as for a variety of other friction kernels, the GLE~\eqref{genLE}, 
which is a complicated integro-differential equation, 
can be replaced by a system of differential equations on a finite dimensional 
extended phase space~\cite{Ferrario79,Grigolini82,Marchesoni83,Martens02}
with an auxiliary coordinate
\begin{equation}
	\label{zetaDef}
	\zeta = -\int_{-\infty}^t \gamma(t-s)\,\dot x(s)\,ds.
\end{equation}
On the extended phase space, the GLE with exponential friction can be represented by 
the system of differential equations
\begin{align}
	\label{eqExp}
	\dot x &= v, \nonumber \\
	\dot v &= -\frac{1}{m}\,\frac{\partial U(x)}{\partial x} + \zeta, \nonumber \\
	\dot \zeta &= -\frac{\gamma_0}{\tau}\,v - \frac{1}{\tau}\,\zeta + \xi_\alpha(t)
\end{align}
now with a white noise source $\xi_\alpha$ that satisfies the 
fluctuation--dissipation theorem
\begin{equation}
	\label{flucdis}
	\avg{\xi_\alpha(t) \xi_\alpha(s)}_\alpha = \frac{2\kT\,\gamma_0}{m\tau^2}\,\delta(t-s).
\end{equation}

In the definition of the auxiliary coordinate~\eqref{zetaDef}, the choice of $-\infty$ 
as the lower limit of integration represents the assumption that the system was prepared 
in the infinite past. 
This assumption is essential to guarantee that the phase space is indeed the 
three-dimensional space with coordinates $x$, $v$ and $\zeta$, 
rather than a submanifold thereof~\cite{Bartsch09}. 
In thermal equilibrium, the auxiliary coordinate follows a 
Gaussian distribution with zero mean and variance
\begin{equation}
	\label{zetaVar}
	\avg{\zeta^2} = \frac{\kT \gamma_0}{m \tau},
\end{equation}
i.e.,~it is not correlated with either position or velocity~\cite{Marchesoni83}.
Accordingly, in the rate calculation, the average over IC's in Eq.~\eqref{Jexact} 
must be supplemented by an average over the distribution of the auxiliary coordinate.

\subsection{Dynamics near a harmonic barrier}

\begin{figure}
\includegraphics{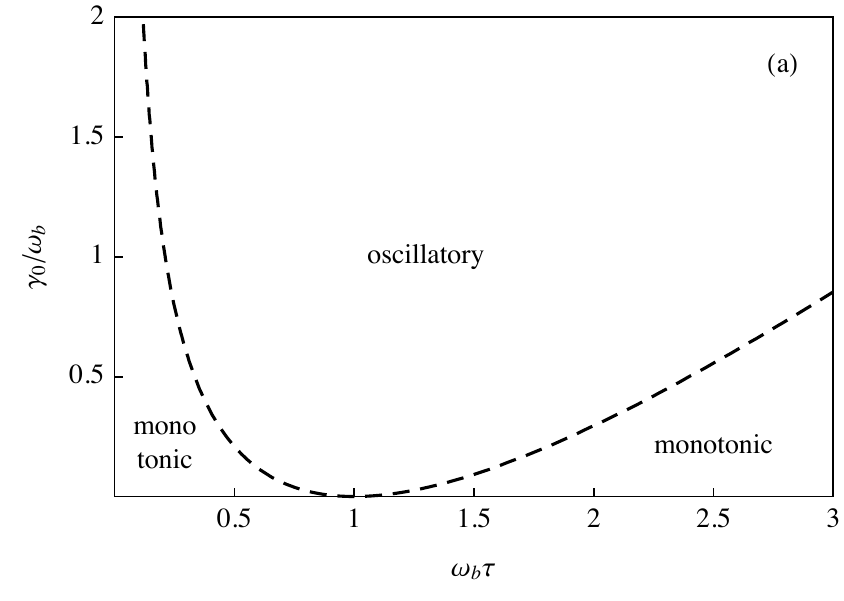}
\medskip
\includegraphics{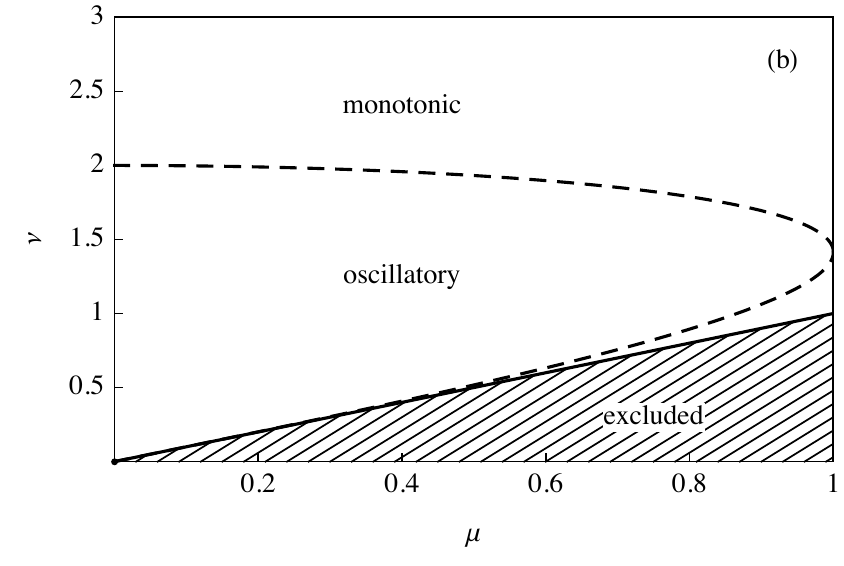}
 \caption{Parameters leading to oscillatory and monotonic behavior for 
  the GLE given by Eq.~\eqref{genLE} with exponential friction. 
  (a) Physical parameters $\gamma_0$ and $\tau$. 
  (b) Parameters~$\mu$ and~$\nu$.}
 \label{fig:1}
\end{figure}

In the harmonic approximation and temporarily neglecting the noise, 
the equations of motion (EoM)~\eqref{eqExp} can be rewritten as
\begin{equation}
        \label{duMu}
	\dot{\vec u} = \mathbf{M} \vec u,
\end{equation}
with the coefficient matrix
\begin{equation}
	\mathbf{M}= \begin{pmatrix}
		0 & 1 & 0 \\
		\omb^2 & 0 & 1 \\
		0 & -\displaystyle \frac{\gamma_0}{\tau} & \displaystyle-\frac{1}{\tau}
	\end{pmatrix},
\end{equation}
and the phase space vector
\[
	\vec u = \begin{pmatrix} x \\ v \\ \zeta \end{pmatrix}.
\]
The eigenvalues $\lambda_0$, $\lambda_1$ and $\lambda_2$ of the matrix~$\mathbf{M}$, 
obtained as the zeros of the characteristic polynomial
\begin{equation}
	\label{charPoly}
	P(\lambda) = -\lambda^3 - \frac{1}{\tau}\,\lambda^2 
		+ \left(\omb^2-\frac{\gamma_0}{\tau}\right) \lambda + \frac{\omb^2}{\tau}.
\end{equation}
are, in general, different.
The corresponding eigenvectors are
\begin{equation}
	\label{evec}
	\vec{\tilde{u}}_i = \begin{pmatrix} 1 \\ \lambda_i \\ \lambda_i^2-\omb^2
 \end{pmatrix}.
\end{equation}
Algebraic expressions for the eigenvalues could in principle be given, 
but they are unwieldy.
More useful are the Vieta relations obeyed by the eigenvalues
\begin{subequations}
\label{Vieta}
\begin{align}
	& \lambda_0 + \lambda_1 + \lambda_2 = -\frac 1{\tau}, \label{Vieta1} \\
	& \lambda_0 \lambda_1 + \lambda_0 \lambda_2 + \lambda_1 \lambda_2 
		= \frac{\gamma_0}{\tau} - \omb^2, \label{Vieta2} \\
	& \lambda_0 \lambda_1 \lambda_2 = \frac{\omb^2}{\tau}. \label{Vieta3}
\end{align}
\end{subequations}
They can be obtained by multiplying out the factorized form
\[
	P(\lambda) = (\lambda_0-\lambda)(\lambda_1-\lambda) (\lambda_2-\lambda)
\]
of the characteristic polynomial and then comparing coefficients.
As
\begin{equation}
        \label{eq:P}
	P(0) = \frac{\omb^2}{\tau}>0 \qquad \text{and} \qquad
	P(\omb) = -\frac{\gamma_0\omb}{\tau}<0,
\end{equation}
at least one of the eigenvalues, say $\lambda_0$, 
must be real and lie between $0$ and $\omb$. 
This eigenvalue describes an unstable direction in phase space. 
The two remaining eigenvalues, $\lambda_1$ and~$\lambda_2$, 
must either be both real and negative, or form a complex conjugate pair 
with real negative parts since, according to the Vieta relations~\eqref{Vieta},
\begin{equation}
      \label{l1+l2_n1l2}
  \lambda_1+\lambda_2 = -\frac{1}{\tau} - \lambda_0 < 0 \quad \text{and}\quad
  \lambda_1\lambda_2 = \frac{\omb^2}{\tau\lambda_0} > 0.
\end{equation}
In either case, a trajectory will approach the origin in the stable 
directions as $t\to\infty$, 
either in an oscillatory manner (if $\lambda_1$ and $\lambda_2$ are complex), 
or monotonically (otherwise). 
The boundary between these different types of behavior in parameter space is 
given by the condition $\lambda_1=\lambda_2$. 
In this case, the discriminant of the characteristic polynomial 
$P(\lambda)$ must be zero. 
The boundary curve obtained in this way is shown in 
Fig.~\ref{fig:1}(a).
Observe that it separates the distinct regions of parameter space
described above.

The corresponding linearized system, $f(x)=0$, is described by three parameters: 
the barrier frequency $\omb$, the damping constant $\gamma_0$, 
and the bath correlation time~$\tau$,
that have either the dimension of a time or an inverse time. 
The three eigenvalues $\lambda_i$ also have the dimension of an inverse time. 
It is convenient to express these quantities in terms of $\omb$, 
which sets the overall time scale, and the two dimensionless parameters
\begin{equation}
	\label{munu}
	\mu = \frac{\lambda_0}{\omb},
	\quad \text{and} \quad
	\nu^2 = \frac{\lambda_0(1+\lambda_0\tau)}{\omb^2\,\tau}
		= \mu^2 \left(1+\frac{1}{\mu\, \omb \tau}\right).
\end{equation}
The parameter $\mu$ takes values between $0$ and $1$, 
while $\nu$ varies between $\mu$ and $\infty$. 
In the white noise limit, $\tau\to 0$ and consequently $\nu\to\infty$. 
This new parameter space is illustrated in Fig.~\ref{fig:1}(b). 
The boundary between monotonic and oscillatory behavior is now
given by the simple condition
\[
	2\mu = \nu \sqrt{4-\nu^2}.
\]

In order to solve Eq.~\eqref{duMu}, we introduce now the diagonal coordinates $z_i$,
by decomposing the phase space vector
\begin{equation}
	\label{diagTrafo}
	\vec u = z_0 \vec{\tilde{u}}_0 + z_1 \vec{\tilde{u}}_1 + z_2 \vec{\tilde{u}}_2
\end{equation}
in the basis set of eigenvectors $\vec{\tilde{u}}_i$. 
In components, the transformation~\eqref{diagTrafo} reads
\begin{align}
	\label{fromDiag}
	x &= z_0 + z_1 + z_2, \nonumber \\
	v &= \lambda_0 z_0 + \lambda_1 z_1 + \lambda_2 z_2, \nonumber \\
	\zeta &= (\lambda_0^2-\omb^2) z_0 + (\lambda_1^2-\omb^2) z_1 
		+ (\lambda_2^2-\omb^2) z_2.
\end{align}
Its inverse is given by
\begin{equation}
	\label{toDiag}
	(\lambda_i-\lambda_j)(\lambda_i-\lambda_k)\,z_i
	= (\lambda_j\lambda_k + \omb^2) x - (\lambda_j+\lambda_k) v + \zeta,
\end{equation}
where the indices $i,j,k=0,1,2$ always take different values. 
In the new coordinates the EoM~\eqref{eqExp} take the form
\begin{equation} \label{eqmotion}
	\dot z_i = \lambda_i z_i 
		+ K_i\,f(x) + \frac{1}{F_i}\,\xi_\alpha(t) ,
\end{equation}
where the abbreviations
\begin{equation*}
	K_i = - \frac{\lambda_j+\lambda_k}{(\lambda_i-\lambda_j)(\lambda_i-\lambda_k)}
\end{equation*}
and
\begin{equation*}
	F_i = (\lambda_i-\lambda_j)(\lambda_i-\lambda_k)
\end{equation*}
have been used.

In the next section, we describe how Eqns.~\eqref{eqmotion} 
can be solved using a perturbative scheme.
%
\section{Time-dependent invariant manifolds}
\label{sec:mf}

The equations~\eqref{eqmotion}, describing the linearized motion of the system, 
can be solved by making the shift of origin in the relative coordinates
\begin{equation} \label{relcoor}
	\Delta z_i(t) = z_i(t) - z_i^\ddag(t), \quad i=0, 1, 2,
\end{equation}
where $z_i^\ddag$ are the components of the TS~trajectory, 
which is defined as
\begin{equation}
	\label{TSDef}
	z_i^\ddag(t) = \frac{1}{F_i}\,
		S[\lambda_i, \xi_\alpha; t] ,
\end{equation}
with the $S$~functionals
\begin{equation}
    \label{SDef}
    S_{t'}[\mu, g;t] = \begin{cases}
            \displaystyle -\int_t^\infty g(t')\,\exp(\mu(t-t')) \,dt' \!\!\!
                & :\; \Re\mu>0, \\[3ex]
            \displaystyle +\int_{-\infty}^t g(t')\,\exp(\mu(t-t')) \,dt' \!\!\!
                & :\; \Re\mu<0.
        \end{cases}
\end{equation}
introduced in Refs.~\onlinecite{Bartsch05b, Kawai07a}.
The Subscript~$t'$ indicates the integration variable, 
and it will be omitted unless necessary to avoid ambiguities.

The TS~trajectory clearly depends on the realization~$\alpha$ of the noise. 
It is the only trajectory that for a given noise sequence remains  
(``jiggling'') in the vicinity of the saddle point for all times. 
By contrast, a typical trajectory will descend either 
into the reactant or product wells in the distant past or  future.
Notice that the TS trajectory defined by Eq.~\eqref{TSDef} is analogous to 
that appearing in the Refs.~\onlinecite{Bartsch05b,Kawai07a,Revuelta12,Bartsch12}.
Since phase space is three-dimensional, it has the additional coordinate $z_2^\ddag(t)$,
which vanishes in the white noise limit, since 
in that case~$\lambda_2 \rightarrow -\infty$.

From the fluctuation-dissipation relation~\eqref{flucdis} for the white noise source $\xi_\alpha$, 
the correlation functions for the components of the TS~trajectory are found to be
\begin{subequations}
\label{correTS}
\begin{align}
    \langle z_0^\ddag (t) z_0^\ddag (0) \rangle_\alpha
  &= \frac{\kT \gamma_0}{m \tau^2\lambda_0 F_0^2}\, e^{-\lambda_0 t}, \label{z0tz00} \\
    \langle z_0^\ddag (t) z_i^\ddag (0) \rangle_\alpha
  &=0, \label{z0tzi0} \\
    \langle z_i^\ddag (t) z_0^\ddag (0) \rangle_\alpha
  &=\frac{2\kT\gamma_0 \left(e^{-\lambda_0 t} 
     - e^{\lambda_i t}\right)}{m \tau^2(\lambda_0+\lambda_i)F_0F_i} \label{zitz00}, \\
    \langle z_i^\ddag (t) z_j^\ddag (0) \rangle_\alpha
  &=-\frac{2\kT\gamma_0}{m \tau^2(\lambda_i+\lambda_j)F_iF_j}\,e^{\lambda_i t}, 
 \label{zitzj0}
\end{align}
\end{subequations}
where $i,j=1,2$ and $t\ge 0$.

The EoM~\eqref{eqmotion} for the relative coordinates~\eqref{relcoor} 
simplify, in this limit, to 
\begin{equation}
	\label{relEq}
	\Delta \dot z_i = \lambda_i \, \Delta z_i + K_i \, f(x),
\end{equation}
which is time-independent. 
Notice that the influence of the stochastic driving does, however, appear
implicitly through the time-dependent shift of origin to the TS~trajectory 
[cf.~Eq.~\eqref{relcoor}]. 

In the harmonic limit, $f(x)=0$, Eqns.~\eqref{relEq} are decoupled and 
can be easily solved as
\begin{equation}
    \label{zjEqIntharm}
    \Delta z_j(t) = \Delta z_j(0) e^{\lambda_j t}, \quad j=0, 1, 2.
\end{equation}

The coordinates $\Delta z_0(t)$ and~$\Delta z_j(t)$ (for $j=1,2$) 
have very different time dependence because $\lambda_0>0$ and $\Re\lambda_j<0$:
$\Delta z_0(t)$ grows exponentially in time, 
whereas $\Delta z_1$ and $\Delta z_2$ shrink.
All trajectories that asymptotically approach the TS~trajectory as $t\to\infty$ 
lie in the plane $\Delta z_0=0$.
This plane is called the stable manifold of the TS~trajectory.
Similarly, trajectories that approach the TS~trajectory backwards in time, 
as $t\to-\infty$, lie on the $\Delta z_0$ coordinate axis, 
i.e., the line $\Delta z_1=\Delta z_2=0$.
This axis is the unstable manifolds of the TS~trajectory.
\begin{figure}
\includegraphics[width=.9\columnwidth]{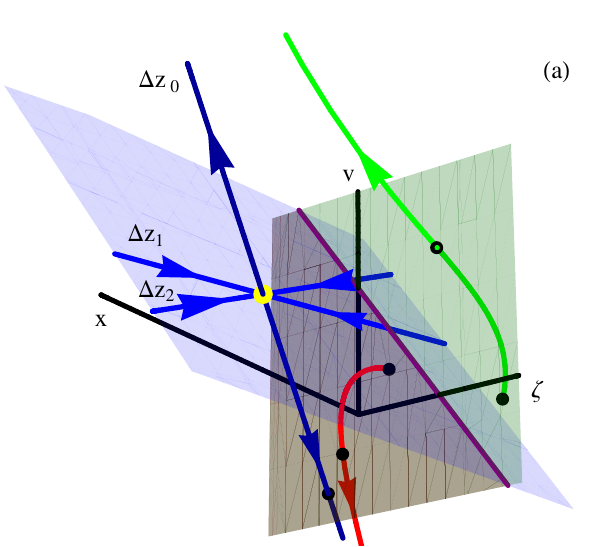}
\includegraphics[width=.9\columnwidth]{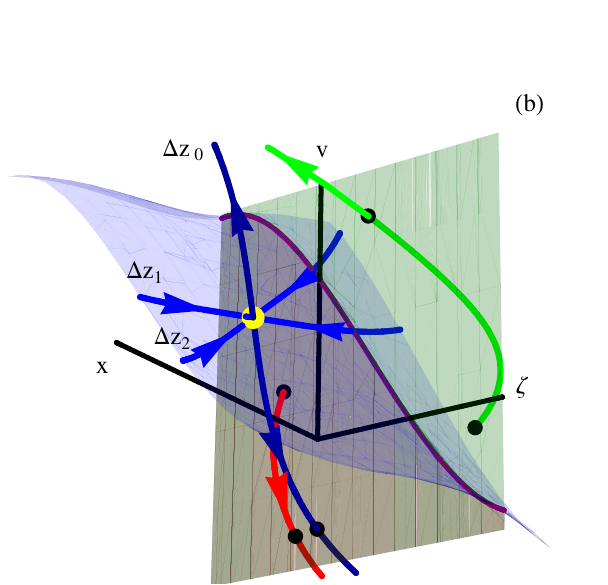}
\caption{Extended phase space of the generalized Langevin equation~\eqref{genLE} 
  for an harmonic (a) and anharmonic (b) potential barrier. 
  Yellow dot: instantaneous position of the TS~trajectory. 
  Dark blue curve: unstable manifold. 
  Light blue surface and trajectories within: stable manifold (SM). 
  The dividing surface ($v$--$\zeta$ plane) is partitioned into 
  reactive (green) and nonreactive (brown) regions by the 
  purple curve that indicates the intersection of the dividing surface 
  with the stable manifold and defines the critical velocity $V^\ddag(\zeta)$. 
  Representative reactive (green) and nonreactive (red) trajectories intersect 
  the dividing surface as indicated by black dots.
}
\label{fig:2}
\end{figure}
The resulting geometry is presented in Fig.~\ref{fig:2}(a) 
where the configuration of these invariant manifolds is shown.
The instantaneous position of the TS~trajectory is indicated by a yellow dot. 
Attached to it there is one unstable direction in which trajectories 
move away from the TS~trajectory, and two stable directions from which 
neighboring trajectories approach the TS~trajectory. 
The plane spanned by the two stable directions is the stable manifold. 
This plot captures the location of the manifolds at 
one instant of time. 
As the TS~trajectory moves, the invariant manifolds will move with it. 
Because the eigenvectors $\tilde{\vec u}_i$ that determine the direction 
of the manifolds do not depend on time, this motion will be rigid, 
without rotation or distortion.

If the barrier potential is not harmonic, i.e.~$f(x)\ne 0$, 
the solution of the EoM~\eqref{relEq} becomes more complicated. 
However, general theorems on dynamical systems guarantee that the invariant 
manifolds persist in the presence of an anharmonic perturbation, 
as long as the perturbation is not too strong. 
They will, however, be distorted
and will no longer be a straight line or a plane,
as shown in Fig.~\ref{fig:2}(b). 
Moreover, the distortion is time-dependent, 
so that the motion of the manifolds is no longer rigid. 
Nevertheless,  the relative configuration 
of the manifolds will qualitatively remain 
as in the harmonic case, 
even though their precise location may be more difficult to determine.

The critical problem in rate theory is the correct identification of 
reactive trajectories.
A careful examination of Fig.~\ref{fig:2} shows that 
a knowledge of the invariant manifolds solves this problem:
All trajectories outside the stable manifold will leave for 
large times the barrier region in the direction of the unstable manifold.
Those that depart in the direction of positive $\Delta z_0$,
which by Eq.~\eqref{fromDiag} corresponds to $x>0$ and therefore 
to the product region, are reactive, 
while those that leave in the direction of negative $\Delta z_0$ 
are nonreactive.
The boundary between these two classes of trajectories is
the stable manifold, which contains all those trajectories 
that approach the TS~trajectory and never leave the barrier region. 

There is an important difference between this scenario and the situation 
in 1-dof systems under the influence of white noise:
In the latter case, the stable and unstable manifolds are both 
one-dimensional curves in a two-dimensional phase space.
Each of them partitions the phase space into two regions, i.e.,
trajectories will enter the reactant or product regions in the distant 
future depending on what side of the stable manifold they are on.
Similarly, the location of a trajectory relative to the unstable manifold 
determines whether it came from reactants or products in the remote past.
The phase space is thus partitioned into four regions with qualitatively 
different behavior.
Under colored noise the stable manifold separates trajectories with 
different future behavior, as we have seen. 
The unstable manifold, by contrast, is only a curve in a three-dimensional 
phase space and is therefore not large enough to partition phase space 
into distinct regions.
This complication, which occurs for white noise only in systems 
with several dof, is introduced by the presence of the auxiliary 
coordinate $\zeta$, that raises the dimension of phase space.

Now we come to an important point.
In order to carry out a rate calculation,
we do not need to classify arbitrary trajectories as reactive or nonreactive.
The flux expression~\eqref{Jexact} contains only trajectories that 
start at the DS~$x=0$.
As shown in Fig.~\ref{fig:2}, the stable manifold intersects the DS in 
the purple curve that can be regarded as the graph of a function $V^\ddag(\zeta)$.
This function will be called the critical velocity.
Though this is not explicit in our notation, the critical velocity 
depends on the realization $\alpha$ of the noise that determines 
the location of the stable manifold.
Trajectories that start with velocities greater than $V^\ddag(\zeta)$ are reactive,
while trajectories with lower initial velocities are not.
This is indicated by the red and green trajectories 
in Fig.~\ref{fig:2},
that start on the DS but on opposite sides of the stable manifold.
The red trajectory begins with an initial velocity smaller 
than $V^\ddag$ and it is therefore nonreactive, while
the green trajectory is reactive because its initial velocity is 
sufficiently high.
Note that both trajectories recross the DS:
The red trajectory starts with positive velocity
and therefore leaves the DS towards the product region,
but then turns around and leaves towards the reactant side.
Conversely, the green trajectory initially moves towards reactants,
but ultimately moves off, towards the product side.
This illustrates why traditional TST,
using the criterion of Eq.~\eqref{chi},
fails in the classification of these two trajectories.

Because the critical velocity encodes all relevant information about reactivity, 
the characteristic function $\chi_r$ can be expressed in terms of it, as
\begin{equation}
  \label{chi_r}
  \chi_r(v_x,\zeta)=\begin{cases}
                		1, & \textnormal{if }  v_x>V^\ddag(\zeta), \\
                		0, & \textnormal{if } v_x<V^\ddag(\zeta).
                    \end{cases}
\end{equation}
Despite its simplicity, Eq.~\eqref{chi_r} is still exact,
and if we carry out the average over initial velocities in the 
flux~\eqref{Jexact} we obtain the transmission factor
\begin{equation}
	\label{defkappaAvg}
	\kappa = \avg{e^{-m V^{\ddag\,2}/2\kT}}_{\alpha,\zeta},
\end{equation}
which must still be averaged over both the noise and the initial value of~$\zeta$. 
Equation~\eqref{defkappaAvg} was originally derived to study the harmonic limit 
in Ref.~\onlinecite{Bartsch08}, but it has also been successfully used 
in Refs.~\onlinecite{Revuelta12,Bartsch12,Revuelta16}
to study anharmonic systems by adequately modifying the harmonic approximation 
to the critical velocity.
In the following section, we describe how this task can be performed 
for colored noise using a perturbative scheme.

\section{The critical velocity}
\label{sec:vCrit}

To calculate the critical velocity $V(\zeta)$ for a fixed value of $\zeta$ in an 
anharmonic potential,
the trajectory on the stable manifold that intersects
the DS $x=0$ at the given value $\zeta$ needs to be determined. 
If we take time~$\protect{t=0}$ as the moment of  intersection, 
we are looking for a trajectory with IC $x(0)=0$ and a given value $\zeta(0)$ 
that is on the stable manifold. 
Its initial velocity $v(0)$ is then the critical velocity $V^\ddag(\zeta(0))$. 

In terms of the diagonal coordinates~\eqref{fromDiag}, the constraints on the  
IC's read
\begin{subequations}
\label{critIC}
\begin{align}
	0 = x(0) &= z_0(0) + z_1(0) + z_2(0) ,  \label{critICx} \\
	\zeta(0) &= \lambda_0^2\, z_0(0) + \lambda_1^2\, z_1(0) + \lambda_2^2\, z_2(0),
		\label{critICzeta}
\end{align}
\end{subequations}
where Eq.~\eqref{critICzeta} has been simplified with the help of Eq.~\eqref{critICx}.
These conditions can be used to eliminate two of the three initial values $z_i(0)$. 
At this point it is convenient to express the initial values 
of the stable dof in terms of the unstable one as
\begin{subequations}
\label{critIC_stab}
\begin{align}
	z_1(0) &= - \frac{\zeta(0) 
		+ (\lambda_2^2-\lambda_0^2)\, z_0(0)}{\lambda_2^2-\lambda_1^2}, \\
	z_2(0) &= + \frac{\zeta(0) 
		+ (\lambda_1^2-\lambda_0^2)\, z_0(0)}{\lambda_2^2-\lambda_1^2}.
\end{align}
\end{subequations}
With these results, the critical velocity is obtained, after some tedious algebra, as
\begin{align}
	\label{V0}
	V^\ddag = v(0) 
		&= \lambda_0\,z_0(0) + \lambda_1\,z_1(0) + \lambda_2\,z_2(0) \nonumber \\
		&= \frac{1}{K_0}\,z_0(0) + \frac{1}{\lambda_1+\lambda_2}\,\zeta(0).
\end{align}
In this expression the value $\zeta(0)$ is known, but $z_0(0)$ is not.
It is determined by imposing the condition
that the trajectory lies on the stable manifold. 

In general, this requires a detailed analysis of the dynamics. 
In the case of a harmonic barrier, however, the stable manifold is simply 
given by $\Delta z_0=0$, or $z_0=z_0^\ddag$, and then 
the critical velocity is
\begin{equation}
	\label{VCritHarm}
	V^{\ddag(0)} = \frac{1}{K_0}\,z^\ddag_0(0) + \frac{1}{\lambda_1+\lambda_2}\,\zeta(0).
\end{equation}

For the anharmonic case, we use 
the perturbative expansion 
\begin{equation}
	\label{VCritExp}
	V^\ddag = V^{\ddag(0)} + \varepsilon V^{\ddag(1)} + \varepsilon^2 V^{\ddag(2)} + \dots
\end{equation}
for the critical velocity, and similar expansions
\begin{equation}
	\label{ICExp}
	z_j(0) = z_j^{(0)} (0) + \varepsilon z_j^{(1)} (0)  + \varepsilon^2 z_j^{(2)} (0) + \dots
\end{equation}
for the coordinates. 
The expansions are carried out under the assumption that the initial value 
$\zeta(0)$ of the auxiliary coordinate remains unchanged. 
Substituting expansions~\eqref{VCritExp} and~\eqref{ICExp} into 
Eqns.~\eqref{critIC_stab} and~\eqref{V0}, one obtains
\begin{align}
	V^{\ddag(k)} &= \frac{1}{K_0}\,\Delta z_0^{(k)}(0), \label{Vk} \\[1ex]
	\Delta z_1^{(k)}(0) &= -\frac {\lambda_2^2-\lambda_0^2}{\lambda_2^2 - \lambda_1^2} \,
					 \Delta z_0^{(k)}(0)
				= K_1 V^{\ddag(k)},    \label{z10} \\[1ex]
\intertext{and}
	\Delta z_2^{(k)}(0) &= +\frac {\lambda_1^2-\lambda_0^2}{\lambda_2^2 - \lambda_1^2} \,
					 \Delta z_0^{(k)}(0)
				= K_2 V^{\ddag(k)}. \label{z20}
\end{align}
Thus, if we can determine the initial value $\Delta z_0^{(k)}(0)$ from the condition 
that the trajectory is on the stable manifold, 
we immediately obtain the corresponding correction to the critical velocity, 
as well as the initial values of the other two coordinates, 
which  in turn determine the trajectories completely.

To proceed, we use a formal solution of the EoM~\eqref{relEq}, 
similar to that used in Refs.~\onlinecite{Revuelta12,Bartsch12}. 
For the unstable coordinate $\Delta z_0$, the general solution is
\begin{equation}
    \label{z0EqInt}
    \Delta z_0(t) = C_0 e^{\lambda_0 t} + 
    	 K_0\,S[\lambda_0,f(x^\ddag+\Delta z_0 + \Delta z_1 + \Delta z_2); t].
\end{equation}
To incorporate the boundary condition that $z_0(t)$ remains bounded as $t\to\infty$, 
$C_0=0$ must be set.

The solutions for the stable coordinates can be written as
\begin{equation}
    \label{zjEqInt}
    \Delta z_j(t) = \Delta z_j(0) e^{\lambda_j t} + 
            K_j\,\bar S[\lambda_j,f(x^\ddag+\Delta z_0 + \Delta z_1 + \Delta z_2); t] 
\end{equation}
for $j=1,2$ in terms of the modified $S$~functional~\cite{Bartsch12}
\begin{equation}
    \label{SmodDef}
    \bar S_{t'}[\mu,g;t] = \int_0^t g(t') e^{\mu(t-t')}\,dt'
\end{equation}
that satisfies the differential equation
\[
  \frac{d}{dt} \bar S[\mu,g;t] = \mu\, \bar S[\mu,g;t] + g(t)
\]
and the IC $\bar S[\mu,g;0]=0$.

The integral equations~\eqref{z0EqInt} and~\eqref{zjEqInt} represent only a formal 
solution to the EoM, since the unknown functions $\Delta z_i(t)$ 
occur on the right hand side. 
They are, however, well suited to a perturbative treatment: for a harmonic barrier, 
i.e.~$f(x)=0$, we obtain the solutions 
$\Delta z_0(t)=0$ and $\Delta z_j(t)=\Delta z_j(0)\,e^{\lambda_jt}$ for $j=1,2$. 
If we can insert this result into Eqns.~\eqref{z0EqInt} and~\eqref{zjEqInt}, 
they will yield first order corrections to the stable manifold. 
This procedure can be iterated to obtain, in principle, corrections of arbitrarily 
high order.

In practice, solving the EoM~\eqref{z0EqInt} and~\eqref{zjEqInt} perturbatively 
requires an expansion not around $x=0$ but around the harmonic trajectory
\begin{equation} \label{X}
	X(t) = x^\ddag(t) + \Delta z_1(0)\,e^{\lambda_1 t} + \Delta z_2(0)\,e^{\lambda_2 t},
\end{equation}
which can be split into a part that depends solely on the realization of the noise
\begin{align} \label{Xalpha}
X_{\alpha}(t)=x^\ddag(t) 
&+\left[\frac{K_{1}}{K_{0}} z_{0}^\ddag(0)-z_{1}^\ddag(0)\right] e^{\lambda_{1} t} 
    \nonumber \\
&+\left[\frac{K_{2}}{K_{0}} z_{0}^\ddag(0)-z_{2}^\ddag(0)\right] e^{\lambda_{2} t},
\end{align}
and another term
\begin{align} \label{Xperp}
   X_{\perp}(t) &= \frac{\zeta(0)}{\lambda_{1}^2-\lambda_{2}^2} \left(e^{\lambda_{1} t} 
                   -  e^{\lambda_{2} t}\right)
\end{align}
that depends on the IC.
Then, the coordinate~$x$ can be expanded as
\begin{equation}
	\label{posExp}
	x(t) = X(t) + \varepsilon \Delta x^{(1)}(t) + \varepsilon^2 \Delta x^{(2)}(t) + \dots,
\end{equation}
where
\[
	\Delta x^{(k)}(t) = \Delta z_0^{(k)}(t) +  \Delta z_1^{(k)}(t) +  \Delta z_2^{(k)}(t)
\]
are the corrections of order $k$ to the position $x$, and
\begin{subequations}
    \label{z0jk}
   \begin{align}
    \Delta z_0^{(k)}(t) &=
     	 K_0\,S[\lambda_0,f^{(k)}; t],     \label{z0k} \\
    \Delta z_j^{(k)}(t) &= \Delta z_j^{(k)}(0) e^{\lambda_j t} 
                           + K_j\,\bar S[\lambda_j,f^{(k)}; t],     
 \label{zjk}
    \end{align}
\end{subequations}
with $j=1, 2$, are the corrections to Eqns.~\eqref{z0EqInt} and~\eqref{zjEqInt}. 
The~$f^{(k)}$ terms appearing in Eq.~\eqref{z0jk} are the coefficients 
in the expansion of the anharmonic force:
\begin{align}
	f(X+\varepsilon \Delta x^{(1)} + \dots ) &=
		 - \varepsilon c_3 X^2  \nonumber \\
  &\quad \, - \varepsilon^2 (2c_3 X\,\Delta x^{(1)} + c_4 X^3) + \dots 
    \nonumber \\
  &= \varepsilon\,f^{(1)} + \varepsilon^2 f^{(2)} + \dots
\end{align}
It is important to note that $f^{(1)}$ is a known function of time because 
it depends solely on the harmonic trajectory~$X$. 
The next correction term $f^{(2)}$ depends on $X$ and the first order 
correction to the position, $\Delta x^{(1)}$. 
Similarly, each $f^{(k)}$ will be known once the lower order corrections 
to position have been evaluated up to order~$k-1$. 

The first order corrections to the relative coordinates can be calculated 
using Eqns.~\eqref{z10},~\eqref{z20} and~\eqref{z0jk} which yield
\begin{subequations}
   \label{z0j1}
   \begin{align}
	\Delta z_0^{(1)}(t) &= K_0\, S[\lambda_0, f^{(1)}; t] \nonumber \\
		&= -K_0 \,c_3\,S[\lambda_0,X^2;t], \label{z01} \\
	\Delta z_j^{(1)}(t) &= K_j\,V^{\ddag(1)} e^{\lambda_j t} 
      + K_j \bar S[\lambda_j, f^{(1)}; t]. \label{zj1}
   \end{align}
\end{subequations}
The first order correction to the critical velocity can then be obtained by 
combining Eqns.~\eqref{Vk} and~\eqref{z01} 
\begin{align}
        \label{V1}
	V^{\ddag(1)} &= -c_3\,S[\lambda_0,X^2;0].
\end{align}
The second order correction to the critical velocity is calculated in a similar way,
which yields
\begin{align}
        \label{V2}
	V^{\ddag(2)} &= -2 c_3\,S[\lambda_0, X\,\Delta x^{(1)}; 0] - c_4\,S[\lambda_0, X^3; 0] ,
\end{align}
with
\begin{align}
	\Delta x^{(1)}(t) &= \Delta z_0^{(1)} (t)+  \Delta z_1^{(1)}(t) 
  +  \Delta z_2^{(1)}(t) \nonumber .
\end{align}

In the next section, we  explain how Eqns.~\eqref{VCritHarm}, 
\eqref{V1} and \eqref{V2} can be used to obtain analytical corrections 
to the transmission factor~\eqref{kappa}.

\section{The transmission factor}
\label{sec:kappa}

The transmission factor~\eqref{kappa} can be expanded in terms of the 
perturbative parameter~$\varepsilon$ by substituting Eq.~\eqref{VCritExp} 
in Eq.~\eqref{defkappaAvg}, this rendering 
\begin{align} 
\label{kappaAvgExpan}
  \kappa &= \kappa^{(0)} + \varepsilon \kappa^{(1)} + \varepsilon^2 \kappa^{(2)} + \ldots,
\end{align}
where
\begin{subequations}
\label{kappaAvg}
\begin{align}
  \kappa^{(0)} &= \avg{ P }_{\alpha\zeta}, \label{kappaAvg0} \\
  \kappa^{(1)} &= -\frac{m}{\kT} \avg{ P V^{\ddag(0)} V^{\ddag(1)}}_{\alpha\zeta}, 
        	\label{kappaAvg1}\\
  \kappa^{(2)} &= \frac{m^2}{2(\kT)^2} \avg{P V^{\ddag(0)\,2} V^{\ddag(1)\,2}}_{\alpha\zeta}  
           \nonumber \\
               &\quad - \frac{m}{\kT} \avg{P V^{\ddag(0)} V^{\ddag(2)}}_{\alpha\zeta}
                - \frac{m}{2\kT} \avg{P V^{\ddag(1)2}}_{\alpha\zeta} , 
          \label{kappaAvg2}
\end{align}
\end{subequations}
with the abbreviation
\begin{equation}
    \label{PDef}
    P = \exp\left(-\frac{m V^{\ddag(0)2}}{2\kT}\right).
\end{equation}
To evaluate Eqns.~\eqref{kappaAvg}, we need to compute averages 
of the form $\langle P (\ldots) \rangle_{\alpha \zeta}$, 
which we will call distorted correlation functions. 
This problem will be addressed in the following subsection.

\subsection{Distorted correlation functions}
\label{sec:distcorrfunc}
The factor~$P$ appearing in Eqns.~\eqref{kappaAvg} can be absorbed into a 
modified covariance matrix. 
This is done similarly to Refs.~\onlinecite{Revuelta12,Bartsch12}, 
where full details of the procedure are given. 
Assume that the random variables $( w_1=V^\ddag_0, w_2, w_3, \dots, w_n)$
follow a multidimensional Gaussian distribution with zero mean and 
covariance matrix $\Sigma$. 
Introduce a modified covariance matrix $\Sigma_0$ that satisfies
\[
    \Sigma_0^{-1} = \Sigma^{-1} + \frac{m}{\kT}\,J ,
\]
with
\[
    J = \begin{pmatrix}
        1 & 0 & 0 & \dots \\
        0 & 0 & 0 & \dots \\
        0 & 0 & 0 & \dots \\
        \vdots & \vdots & \vdots & \ddots
    \end{pmatrix}.
\]
Using $\avg{...}_0$ to denote an average over a multidimensional
Gaussian distribution with zero mean and covariance matrix $\Sigma_0$, 
we can write
\begin{align}
\avg {P (\dots)}_{\alpha\zeta}
 = \frac{\lambda_0}{\omb}\, \avg{...}_0.
    \label{modGaussian}
\end{align}

The matrix $\Sigma_0$ is explicitly given by
\begin{align}
   \label{S0}
    \Sigma_0 &= \Sigma-\frac{m}{\kT+m\sigma^2}\,\Sigma J \Sigma,
\end{align}
where~$\sigma^2=\avg{V^{\ddag(0)\,2}}_{\alpha\zeta}$. 
Moreover,~$\sigma^2$ can be easily computed by noting that the harmonic 
approximation to the critical velocity~\eqref{VCritHarm} is a sum of two Gaussian 
random variables that are independent because the first term,~$z_0^\ddag(0)$, 
depends only on the noise and the second,~$\zeta(0)$, only on the IC. 
With the help of Eqns.~\eqref{zetaVar} and~\eqref{z0tz00} we can compute
\begin{eqnarray}
	\label{sigmaV0}
	\sigma^2 &=& \avg{V^{\ddag(0)\,2}}_{\alpha\zeta} \nonumber \\ 
		     &=& 
		\frac{(\lambda_0-\lambda_1)^2(\lambda_0-\lambda_2)^2}
			{(\lambda_1+\lambda_2)^2}\, \avg{z_0^{\ddag\,2}(0)}_\alpha  \nonumber \\ 
		     &\quad 
            &+ \frac{1}{(\lambda_1+\lambda_2)^2}\,\avg{\zeta^2(0)}_\zeta \nonumber \\ 
	        &=& \frac{\kT\gamma_0}{m\lambda_0(1+\lambda_0\tau)},
\end{eqnarray}
In the last step it has been taken into account that 
$(\lambda_1+\lambda_2)\tau = -(1+\lambda_0\tau)$ according to~\eqref{Vieta}.
The modified covariance matrix~\eqref{S0} can then be simplified to
\begin{align}
	\label{modSigma}
	\Sigma_0 &= \Sigma - \frac{m}{\kT}\,\frac{\lambda_0^2}{\omb^2}\,\Sigma J\Sigma
\end{align}
with the help of the algebraic relation
\begin{align}
	\label{GHDivision}
	\frac{\lambda_0^2}{\omb^2}\left[\lambda_0(1+\lambda_0\tau)+\gamma_0\right]
		&= -\frac{\tau}{\omb^2}\,\lambda_0\,P(\lambda_0) + \lambda_0(1+\lambda_0\tau)
			\nonumber \\
		&= \lambda_0(1+\lambda_0\tau).
\end{align}
For its components we find
\begin{equation}
    \label{modSigmaComp}
    \avg{w_i w_j}_0 = \avg{w_i w_j}_{\alpha\zeta}
        - \frac{m}{\kT}\,\frac{\lambda_0^2}{\omb^2}
        	\avg{V^{\ddag(0)} w_i}_{\alpha\zeta} \! \! \! \avg{V^{\ddag(0)} w_j}_{\alpha\zeta} \! ,
\end{equation}
which allows one to obtain the moments of the distorted Gaussian distribution, 
once those of the original Gaussian are known. 
If we take a random variable $w_\alpha$ depending only on the noise, 
and another one $w_\zeta$ depending only on the IC, 
the original covariance $\avg{w_\alpha w_\zeta}_{\alpha\zeta}$ vanishes, 
but $\avg{w_\alpha w_\zeta}_0$ is, in general, nonzero, since $V^{\ddag (0)}$ 
depends both on the noise and on the IC $\zeta(0)$.

If $w_i=V^{\ddag (0)}$, the modified moment becomes a multiple of the original one 
\begin{align}
	\avg{V^{\ddag(0)} w_j}_0 &=
		\avg{V^{\ddag(0)} w_j}_{\alpha\zeta} 
			\left( 1-\frac{m}{\kT}\,\frac{\lambda_0^2}{\omb^2} 
				\avg{V^{\ddag(0)\,2}}_{\alpha\zeta} \right) \nonumber \\[1ex]
		&= \avg{V^{\ddag(0)} w_j}_{\alpha\zeta}
			\,\frac{\omb^2(1+\lambda_0\tau)-\lambda_0\gamma_0}
				{\omb^2(1+\lambda_0\tau)}
			\nonumber\\[1ex]
		&= \frac{\lambda_0^2}{\omb^2}\avg{V^{\ddag(0)} w_j}_{\alpha\zeta}. \label{V0wj}
\end{align}
In this calculation we have used Eq.~\eqref{sigmaV0} and the fact that
\[
	\lambda_0^2(1+\lambda_0\tau) = (\omb^2\tau-\gamma_0)\lambda_0 + \omb^2
\]
because $\lambda_0$ is a zero of the characteristic polynomial~\eqref{charPoly}.
In particular, we have
\begin{align}
	\avg{V^{\ddag(0)\,2}}_0 
		&= \frac{\lambda_0^2}{\omb^2}\,\avg{V^{\ddag(0)\,2}}_{\alpha\zeta}= 
          \frac{\kT}{m}\left(1-\frac{\lambda_0^2}{\omb^2}\right)
	\label{VVcorr}
\end{align}
since by a similar argument
\[
	\lambda_0\gamma_0 = -\lambda_0^3\tau - \lambda_0^2 + \omb^2\lambda_0\tau+\omb^2
		= (1+\lambda_0\tau)(\omb^2-\lambda_0^2).
\]

As will be seen in Sec.~\ref{subsec:results1d}, the calculation of reaction 
rates requires the correlation functions
\begin{eqnarray}
	\label{VXcorr}
	\avg{V^{\ddag(0)}\,X(t)}_0
	&=& \frac{\kT}{m \, \lambda_0}\left[e^{-\lambda_0 t}
		+\frac{\lambda_2\tau\,(\lambda_0+\lambda_2)}
			{(\lambda_2-\lambda_1)}\,e^{\lambda_1t} \right.\nonumber \\
		& &\qquad \qquad + \left.\frac{\lambda_1\tau\,(\lambda_0+\lambda_1)}
			{(\lambda_1-\lambda_2)}\,e^{\lambda_2t} \right]
\end{eqnarray}
and
\begin{align}
	\label{XXcorr}
	\avg{X(t)\,X(s)}_0 &= 
                  \frac{\kT}{m}\left[\frac{K_0}{\lambda_0}\,e^{-(t-s)\lambda_0}
		+\frac{K_1}{\lambda_1}\,e^{(t-s)\lambda_1} \right. \nonumber \\[1ex]
		&+\frac{K_2}{\lambda_2}\,e^{(t-s)\lambda_2}
                 +\frac{1}{\omb^2}\,e^{-(t+s)\lambda_0} \nonumber \\[1ex]
		&+\frac{\lambda_0 \lambda_2 +\omb^2}{F_1 \omb^2}\,
			\left(e^{-\lambda_0t+\lambda_1 s} + e^{-\lambda_0 s + \lambda_1 t}\right) 
			\nonumber\\[1ex]
	& 
		\left.+\frac{\lambda_0 \lambda_1 +\omb^2}{F_2 \omb^2}\,
			\left(e^{-\lambda_0t+\lambda_2 s} + e^{-\lambda_0 s + \lambda_2 t}\right) \right]
\end{align}
for $t\ge s \ge 0$.

Distorted averages involving more than two factors 
of~$V^{\ddag(0)}$ and~$X(t)$ 
can be reduced to the correlation functions~\eqref{VVcorr}, \eqref{VXcorr} 
and~\eqref{XXcorr} by Isserlis' theorem~\cite{Isserlis16,Isserlis18}, e.g.
\begin{align*}
    \avg{w_1 w_2 w_3 w_4}_0 &= \avg{w_1 w_2}_0 \avg{w_3w_4}_0
        + \avg{w_1 w_3}_0 \avg{w_2w_4}_0 \\
      &\,\, + \avg{w_1 w_4}_0 \avg{w_2 w_3}_0.
\end{align*}
This expression contains a sum over all possible pairings of the four factors.
Other even order moments can be evaluated in a similar way,
and all odd order moments are zero.
In this way, the modified averages of arbitrary polynomials can be calculated.

\subsection{Results for the one--dimensional potential}
\label{subsec:results1d}
The correlation functions~\eqref{VXcorr} and~\eqref{XXcorr} allow us to 
evaluate the corrections to the transmission factor. 
For the leading order, Eqns.~\eqref{kappaAvg0} and~\eqref{modGaussian} 
immediately give
\begin{equation}
        \label{kappaGH}
	\kappa^{(0)} = \frac{\lambda_0}{\omb} \avg{1}_0 = \frac{\lambda_0}{\omb},
\end{equation}
which is the well-known Grote--Hynes~\cite{Grote80} result for a harmonic barrier.

The first order rate correction to Eq.~\eqref{kappaGH} can be rewritten as
\begin{align}
	\kappa^{(1)} 
     &= \frac{m}{\kT}\,\frac{\lambda_0}{\omb} \avg{V^{\ddag(0)} V^{\ddag(1)}}_0 
          \nonumber \\
	 &= -\frac{m c_3}{\kT} \,\frac{\lambda_0}{\omb}\, 
              S[\lambda_0, \avg{V^{\ddag(0)}\,X^2}_0 ; 0]
			\nonumber \\
	 &= 0,
\end{align}
which is zero because the correlation function is of third order in 
$V^{\ddag(0)}$ and $X$. 
Similarly, all higher rate corrections of odd order vanish. 
As a consequence, the expansion~\eqref{kappaAvgExpan} contains only even 
powers of the perturbation parameter~$\varepsilon$. 
It will therefore yield an expansion in integer powers of $\kT$, 
rather than an expansion in powers~$\sqrt{\kT}$, 
as one might expect at first sight.

The second order correction in Eq.~\eqref{kappaAvg2} to the rates has three terms. 
The first one is given by
\begin{align}
	&\qquad 
	\frac{m^2}{2(\kT)^2}\,\frac{\lambda_0}{\omb} \avg{V^{\ddag(0)\,2} V^{\ddag(1)\,2}}_0 
      \nonumber \\[1ex]
	&= \frac{m^2 c_3^2}{2(\kT)^2}\,\frac{\lambda_0}{\omb}
		S_t\Big[ \lambda_0, S_s[\lambda_0, \avg{V^{\ddag(0)\,2} 
        X^2(t) X^2(s)}_0 ; 0 ] ; 0 \Big].
\end{align}
The remaining correlation function can be reduced to~\eqref{VXcorr} and~\eqref{XXcorr} 
by means of Isserlis' theorem. 
It will yield a sum of exponentially decaying terms, for which the $S$~functionals, 
which are short-hand notation for the integral~\eqref{SDef}, can be computed. 
The calculation is straightforward with the help of a computer algebra system, 
\textit{Mathematica}~\cite{Mathematica} in our case,
giving the rate correction
\begin{widetext}
\begin{align}
\label{eq:kappa2}
\kappa^{(2)} &= -\frac{c_3^2\, \kT}{6 m \omb^6}
	\frac{ \mu  \left(\mu ^2-1\right)^2}
	{\left(\mu ^2+\nu ^2\right) \left[\mu ^4+2 \mu ^2 \left(\nu ^2-2\right)+4
  		 \nu ^2\right] \left[\mu ^4+\mu ^2 \left(\nu ^2-1\right)+\nu ^2\right]^2 \left[4
   		\mu ^4+\mu ^2 \left(2 \nu ^2-1\right)+\nu ^2\right]} \times \nonumber \\[1ex]
   	& \qquad\qquad \bigg[	2 \left(10 \mu^4+41 \mu ^2+10\right) \nu^{10}+
	 	\left(110 \mu ^4+329 \mu ^2-12\right) \mu^2\nu ^8+ 
		2 \left(115 \mu ^4+197 \mu ^2-28\right) \mu ^4 \nu ^6+ \nonumber\\[1ex]
	& \qquad\qquad\quad 2 \left(115 \mu ^4+22 \mu ^2+8\right) \mu^6 \nu ^4+ 
		2 \left(55 \mu ^4-94 \mu ^2+6\right) \mu^8 \nu^2+
		5 \left(4 \mu^4-17 \mu ^2+4\right) \mu ^{10}
		\bigg] \nonumber\\
	& \quad -\frac{3 \,c_4 \,\kT}{4 m \, \omb^4}
   		\frac{\mu  \left(\mu ^2-1\right)^2 \left(\mu ^2+\nu ^2\right)^2}
   			{ \left[\mu ^4+\mu ^2 \left(\nu ^2-1\right)+\nu ^2\right]^2}.
\end{align}
\end{widetext}
In the limit $\nu\to\infty$, which corresponds to white noise, this expression 
reduces to
\begin{align}
        \label{eq:kappa2white}
	\kappa^{(2)}(\nu\to\infty) &= -\frac{c_3^2\, \kT}{6m\omb^6}\,
	\frac{ \mu  \left(1-\mu ^2\right)^2 }{(1+\mu^2)^2}\,
	\frac{(10+41\mu^2+10\mu^4)}{(2+5\mu^2+2\mu^4)} \nonumber \\
     &\quad -\frac{3 \,c_4 \,\kT}{4 m \omb^4}\,
   \frac{ \mu  \left(1-\mu ^2\right)^2 }
   	{(1+\mu^2)^2},
\end{align}
which agrees with the known result for this case,
see Refs.~\onlinecite{Talkner93,Pollak93a,Revuelta12,Bartsch12}.

\begin{figure}
\centerline{\includegraphics{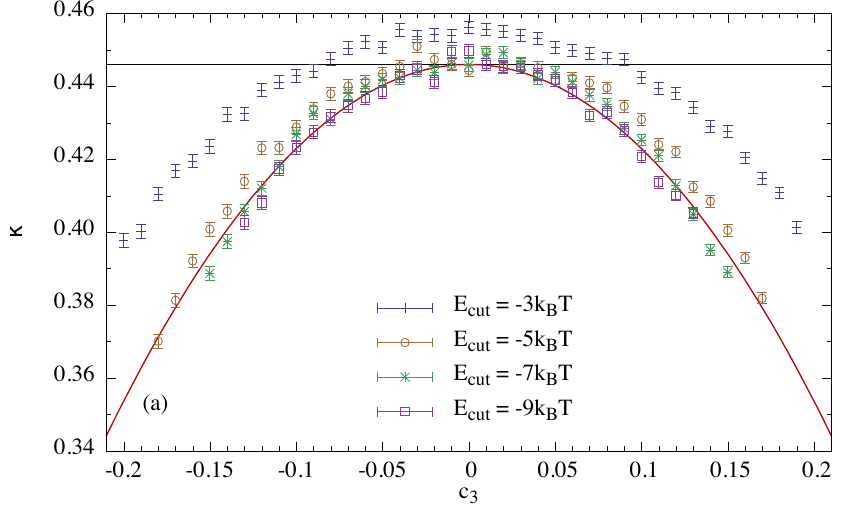}}
\medskip

\centerline{\includegraphics{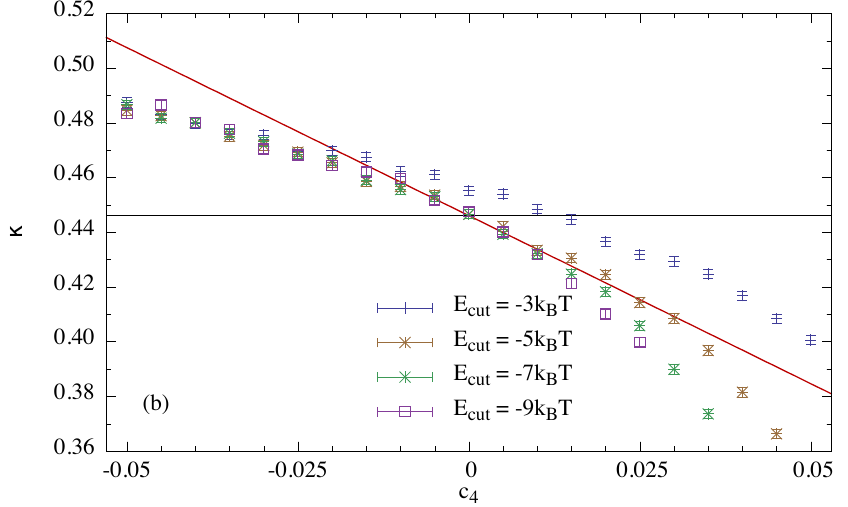}}
\caption{Transmission factor for an anharmonic barrier with 
  (a) cubic and (b) quartic perturbation. 
  Horizontal black line: Grote-Hynes transmission factor given by Eq.~\eqref{kappaGH}, 
  red curve: leading order perturbation theory result obtained 
  from~\eqref{kappaGH}+\eqref{eq:kappa2}. 
  Symbols: numerical simulation with $1\sigma$ statistical error bars and cutoff energy 
  equal to $-3\kT$ (blue plus symbols), $-5\kT$ (brown circles), $-7\kT$ (green stars), $-9\kT$ (purple squares). 
  Parameters used are equal to $m=1$, $\omb=1$, $\gamma_0=5$, $\tau=4$, $\kT=1$.
  }
\label{fig:rates}
\end{figure}

The rate correction obtained from Eq.~\eqref{eq:kappa2} is compared to the 
result of numerical simulations, computed as described in Sect.~\ref{sec:rate}, 
in Fig.~\ref{fig:rates}.
To obtain converged results, the cutoff energy at which trajectories are 
considered to be thermalized on either reactant or product side
must be chosen sufficiently low.
Actually, it must be significantly lower than what would be
required for a similar computation with white noise. 
This effect can be clearly seen in the top panel of the figure.
An energy cutoff of~$-3\kT$ (blue plus symbols) is not enough to identify the reactive 
trajectories reliably, even in the harmonic limit, $c_3=0$,
where the Grote--Hynes result~\eqref{kappaGH} is exact (black horizontal line).
This is due to the memory effect inherent in correlated noise:
The friction force  remembers that the trajectory came from the barrier top
and  therefore tends to push it back up
Thus, a lower energy cutoff has to be chosen.
As can be seen in the same panel,
the numerical simulations for~$-5\kT$ (brown circles), $-7\kT$ (green stars)
and $-9\kT$ (purple squares) provide more accurate results.
Indeed, results are well converged for a cutoff energy of $-7\kT$,
and this value will be used in all further calculations.
The converged transmission factors are in good agreement with the perturbative results.
If $c_3\ne 0$, the potential has a minimum on one side of the barrier.
The cutoff energy cannot be chosen below the minimum, or conversely, 
for given cutoff energy the coupling strength $c_3$ must be chosen such as 
to produce a sufficiently low minimum.
For this reason, the data in the figure cover a smaller range of $c_3$ 
if the cutoff is lower.

Similar comments apply to Fig.~\ref{fig:rates}(b),
where the transmission factor is shown as a function of a quartic 
coupling strength $c_4$.
Again, the numerical results decrease as the cutoff energy is lowered,
finally converging to a limit that is in good agreement with the 
perturbative results if the coupling is not too strong.

\begin{figure}
\includegraphics[]{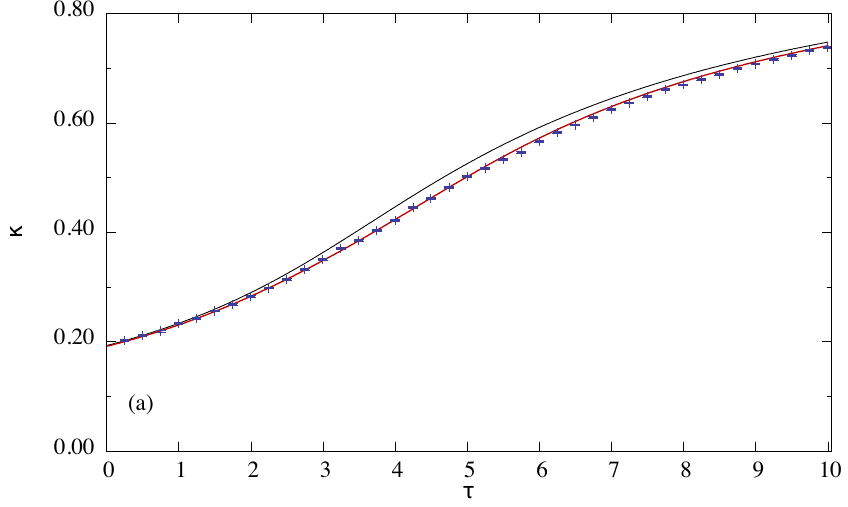}
\medskip
\includegraphics[]{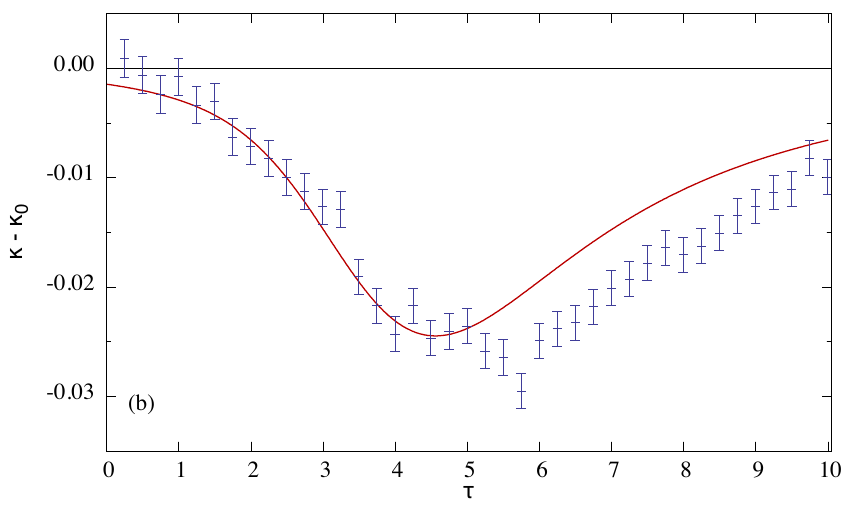}
\caption{Transmission factor as a function of memory time. 
  Top black line: Grote-Hynes transmission factor~\eqref{kappaGH}, 
  bottom red line: leading order perturbation theory obtained 
    from~\eqref{kappaGH}+\eqref{eq:kappa2}. 
  Blue symbols: numerical simulation with 1$\sigma$ error bars and cutoff energy $-7\kT$. 
  (a) Transmission factor, (b) deviation from the harmonic approximation. 
  Parameters used are $m=1$, $\omb=1$, $\gamma_0=5$, $c_3=0.1$, $c_4=0$, $\kT=1$.}
\label{fig:ratesTau}
\end{figure}

Figure~\ref{fig:ratesTau} shows the dependence of the transmission factor 
on the memory time.
This dependence is strong, and more importantly it is largely accounted for
by the harmonic approximation.
Nevertheless, the deviation from the harmonic approximation also varies 
strongly with the memory time.
The absolute value of the anharmonic correction is smallest in the white noise 
limit $\tau\to 0$.
It grows for nonzero memory times, has a maximum 
at~$\tau_\text{min}\approx 4.5$ 
and then it decreases again.
This behavior is qualitatively well described by the leading order perturbation theory.
The agreement between perturbation theory and simulation is excellent also in 
quantitative terms for memory times shorter than $\tau_\text{min}$.
For larger times, it is only approximate.

\begin{figure}
\includegraphics[]{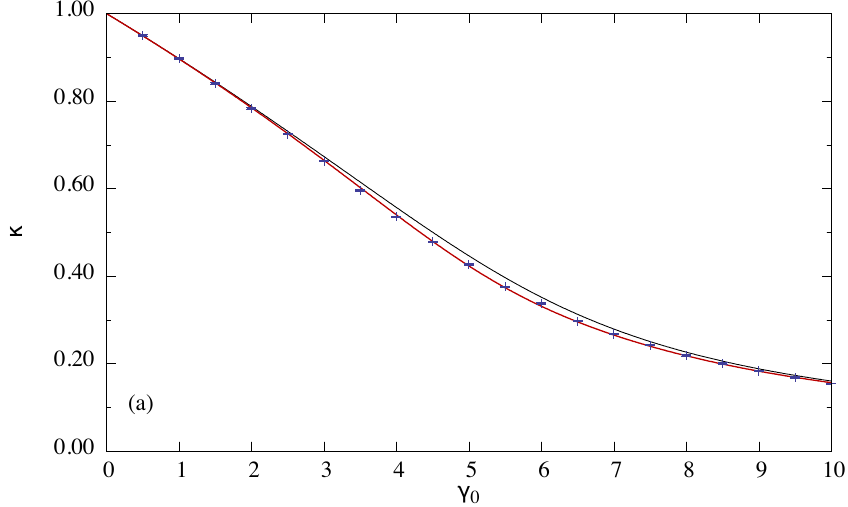}
\medskip
\includegraphics[]{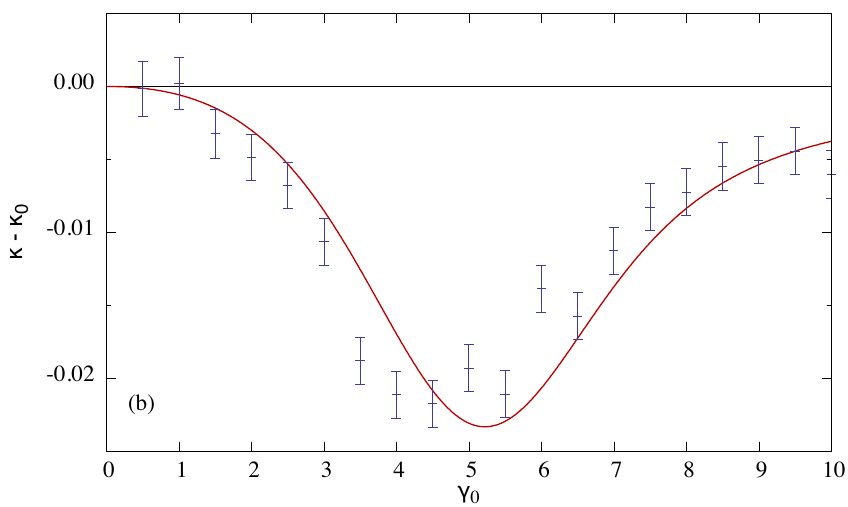}
\caption{Transmission factor as a function of damping. 
  Top black line: Grote-Hynes transmission factor~\eqref{kappaGH}, 
  bottom red line: leading order perturbation theory obtained 
  from~\eqref{kappaGH}+\eqref{eq:kappa2}. 
  Blue symbols: numerical simulation with 1$\sigma$ error bars and cutoff 
  energy $-7\kT$. 
  (a) Transmission factor, 
  (b) deviation from the harmonic approximation. 
  Parameters used are  $m=1$, $\omb=1$, $\tau=4$, $c_3=0.1$, $c_4=0$, $\kT=1$.
}
\label{fig:ratesGamma}
\end{figure}

Similar results are shown in Fig.~\ref{fig:ratesGamma} for the transmission 
factor as a function of damping strength.
The transmission factor depends strongly on the strength of the damping, 
and again most of this dependence is accounted for by the harmonic approximation.
The anharmonic correction is zero for $\gamma_0=0$, increases in magnitude for
nonzero friction, then goes through a minimum and finally  decreases again.
Perturbation theory is in good agreement with the numerical results over 
the entire range of $\gamma_0$.

It should be kept in mind that the results are not physically meaningful 
in the limit of weak damping, because the rate theory outlined in 
Sec.~\ref{sec:rate} assumes that the rate is determined by spatial diffusion.
As $\gamma_0\to 0$, a turnover to an energy diffusion limited rate will occur 
at a value of $\gamma_0$ that depends on the details of the potential well.
It cannot therefore be stated in general how strong the damping has to be for the 
results of Fig.~\ref{fig:ratesGamma} to be applicable. 
For an assessment of the perturbative results, however, this question is not relevant.

Finally, let us remark that the theory outlined here has been also successfully 
applied to more realistic chemical models~\cite{Revuelta16}.
Actually, we were able to accurately reproduce the reaction rates of the
LiNC$\rightleftharpoons$LiCN isomerization reaction in the presence 
of an argon bath computed obtained using all-atom molecular dynamics by using a 
simple one-dimensional model defined along the minimum energy path of the molecule.

\section{Concluding remarks}
\label{sec:conclu}

The computation of chemical reaction rates using TST and similar 
approaches is very common in the chemistry community. 
However, the results rendered by standard
TST depend dramatically on the choice of an adequate DS. 
This is particularly important in reactions that take place in a solvent, 
where a typical reactive trajectory recrosses the DS many times and, 
as a consequence, standard TST grossly overestimates 
the true reaction rate.

In this paper we present a method that overcomes the recrossing problem.
It identifies reactive trajectories precisely by computing the geometrical 
structures that divide the phase space into reactive and nonreactive parts. 
More specifically, all the information on the reactivity of the system 
is encoded in the stable manifold, whose intersection with the DS defines 
a critical velocity that trajectories must exceed in order to be reactive. 
Notice that this procedure is independent of the selected DS as the stable 
manifold acts as a separatrix throughout the (extended) phase space. 
The intersection of the stable manifold with a different DS renders a 
different critical velocity but if a trajectory is reactive it will cross 
each DS with a velocity larger than the corresponding critical velocity.

The method reported here is based on a perturbative scheme.
It extends a previous series of studies%
\cite{Bartsch05b,Bartsch05c,Bartsch06a,Bartsch08,Hernandez10,%
Revuelta12,Bartsch12,Revuelta16} to the case of colored noise and
it has also been successfully applied to calculate the reaction rates
of a realistic molecular system~\cite{Revuelta16}.
Furthermore, it has  enabled us to obtain analytic corrections to the 
Grote--Hynes expression for anharmonic multidimensional potentials,
while providing at the same time a clear geometrical picture 
of the reaction mechanism.

\section*{Acknowledgements}
The research leading these results has received funding from
the Minis\-terio de Econom\'ia y Competitividad under Contract 
MTM2015-63914-P, ICMAT Severo Ochoa under 
SEV-2015-0554, 
and the People Programme (Marie Curie Actions) of the European Union's Seventh 
Framework Programme FP7/2007-2013/under REA Grant Agreement No. 294974.


\begin{thebibliography}{0}%
\makeatletter
\providecommand \@ifxundefined [1]{%
 \@ifx{#1\undefined}
}%
\providecommand \@ifnum [1]{%
 \ifnum #1\expandafter \@firstoftwo
 \else \expandafter \@secondoftwo
 \fi
}%
\providecommand \@ifx [1]{%
 \ifx #1\expandafter \@firstoftwo
 \else \expandafter \@secondoftwo
 \fi
}%
\providecommand \natexlab [1]{#1}%
\providecommand \enquote  [1]{``#1''}%
\providecommand \bibnamefont  [1]{#1}%
\providecommand \bibfnamefont [1]{#1}%
\providecommand \citenamefont [1]{#1}%
\providecommand \href@noop [0]{\@secondoftwo}%
\providecommand \href [0]{\begingroup \@sanitize@url \@href}%
\providecommand \@href[1]{\@@startlink{#1}\@@href}%
\providecommand \@@href[1]{\endgroup#1\@@endlink}%
\providecommand \@sanitize@url [0]{\catcode `\\12\catcode `\$12\catcode
  `\&12\catcode `\#12\catcode `\^12\catcode `\_12\catcode `\%12\relax}%
\providecommand \@@startlink[1]{}%
\providecommand \@@endlink[0]{}%
\providecommand \url  [0]{\begingroup\@sanitize@url \@url }%
\providecommand \@url [1]{\endgroup\@href {#1}{\urlprefix }}%
\providecommand \urlprefix  [0]{URL }%
\providecommand \Eprint [0]{\href }%
\providecommand \doibase [0]{http://dx.doi.org/}%
\providecommand \selectlanguage [0]{\@gobble}%
\providecommand \bibinfo  [0]{\@secondoftwo}%
\providecommand \bibfield  [0]{\@secondoftwo}%
\providecommand \translation [1]{[#1]}%
\providecommand \BibitemOpen [0]{}%
\providecommand \bibitemStop [0]{}%
\providecommand \bibitemNoStop [0]{.\EOS\space}%
\providecommand \EOS [0]{\spacefactor3000\relax}%
\providecommand \BibitemShut  [1]{\csname bibitem#1\endcsname}%
\let\auto@bib@innerbib\@empty
\end{thebibliography}%


\begin{thebibliography}{54}%
\makeatletter
\providecommand \@ifxundefined [1]{%
 \@ifx{#1\undefined}
}%
\providecommand \@ifnum [1]{%
 \ifnum #1\expandafter \@firstoftwo
 \else \expandafter \@secondoftwo
 \fi
}%
\providecommand \@ifx [1]{%
 \ifx #1\expandafter \@firstoftwo
 \else \expandafter \@secondoftwo
 \fi
}%
\providecommand \natexlab [1]{#1}%
\providecommand \enquote  [1]{``#1''}%
\providecommand \bibnamefont  [1]{#1}%
\providecommand \bibfnamefont [1]{#1}%
\providecommand \citenamefont [1]{#1}%
\providecommand \href@noop [0]{\@secondoftwo}%
\providecommand \href [0]{\begingroup \@sanitize@url \@href}%
\providecommand \@href[1]{\@@startlink{#1}\@@href}%
\providecommand \@@href[1]{\endgroup#1\@@endlink}%
\providecommand \@sanitize@url [0]{\catcode `\\12\catcode `\$12\catcode
  `\&12\catcode `\#12\catcode `\^12\catcode `\_12\catcode `\%12\relax}%
\providecommand \@@startlink[1]{}%
\providecommand \@@endlink[0]{}%
\providecommand \url  [0]{\begingroup\@sanitize@url \@url }%
\providecommand \@url [1]{\endgroup\@href {#1}{\urlprefix }}%


\providecommand \urlprefix  [0]{URL }%
\providecommand \Eprint [0]{\href }%
\providecommand \doibase [0]{http://dx.doi.org/}%
\providecommand \selectlanguage [0]{\@gobble}%
\providecommand \bibinfo  [0]{\@secondoftwo}%
\providecommand \bibfield  [0]{\@secondoftwo}%
\providecommand \translation [1]{[#1]}%
\providecommand \BibitemOpen [0]{}%
\providecommand \bibitemStop [0]{}%
\providecommand \bibitemNoStop [0]{.\EOS\space}%
\providecommand \EOS [0]{\spacefactor3000\relax}%
\providecommand \BibitemShut  [1]{\csname bibitem#1\endcsname}%
\let\auto@bib@innerbib\@empty
\bibitem [{\citenamefont {Marcelin}(1915)}]{Marcelin15}%
  \BibitemOpen
  \bibfield  {author} {\bibinfo {author} {\bibfnamefont {R.}~\bibnamefont
  {Marcelin}},\ }\href@noop {} {\bibfield  {journal} {\bibinfo  {journal}
  {Annales de Physique}\ }\textbf {\bibinfo {volume} {3}},\ \bibinfo {pages}
  {120} (\bibinfo {year} {1915})}\BibitemShut {NoStop}%
\bibitem [{\citenamefont {Evans}\ and\ \citenamefont
  {Polanyi}(1935)}]{Evans35}%
  \BibitemOpen
  \bibfield  {author} {\bibinfo {author} {\bibfnamefont {M.~G.}\ \bibnamefont
  {Evans}}\ and\ \bibinfo {author} {\bibfnamefont {M.}~\bibnamefont
  {Polanyi}},\ }\href {\doibase 10.1039/TF9353100875} {\bibfield  {journal}
  {\bibinfo  {journal} {Trans.\ Faraday Soc.}\ }\textbf {\bibinfo {volume}
  {31}},\ \bibinfo {pages} {875} (\bibinfo {year} {1935})}\BibitemShut
  {NoStop}%
\bibitem [{\citenamefont {Wigner}(1938)}]{Wigner38}%
  \BibitemOpen
  \bibfield  {author} {\bibinfo {author} {\bibfnamefont {E.}~\bibnamefont
  {Wigner}},\ }\href@noop {} {\bibfield  {journal} {\bibinfo  {journal} {FDCS}\
  }\textbf {\bibinfo {volume} {34}},\ \bibinfo {pages} {29} (\bibinfo {year}
  {1938})}\BibitemShut {NoStop}%
\bibitem [{\citenamefont {Horiuti}(1938)}]{Horiuti38}%
  \BibitemOpen
  \bibfield  {author} {\bibinfo {author} {\bibfnamefont {J.}~\bibnamefont
  {Horiuti}},\ }\href {\doibase 10.1246/bcsj.13.210} {\bibfield  {journal}
  {\bibinfo  {journal} {Bull. Chem. Soc. Japan}\ }\textbf {\bibinfo {volume}
  {13}},\ \bibinfo {pages} {210} (\bibinfo {year} {1938})}\BibitemShut
  {NoStop}%
\bibitem [{\citenamefont {Toller}\ \emph {et~al.}(1985)\citenamefont {Toller},
  \citenamefont {Jacucci}, \citenamefont {DeLorenzi},\ and\ \citenamefont
  {Flynn}}]{Toller85}%
  \BibitemOpen
  \bibfield  {author} {\bibinfo {author} {\bibfnamefont {M.}~\bibnamefont
  {Toller}}, \bibinfo {author} {\bibfnamefont {G.}~\bibnamefont {Jacucci}},
  \bibinfo {author} {\bibfnamefont {G.}~\bibnamefont {DeLorenzi}}, \ and\
  \bibinfo {author} {\bibfnamefont {C.~P.}\ \bibnamefont {Flynn}},\ }\href@noop
  {} {\bibfield  {journal} {\bibinfo  {journal} {Phys. Rev. B}\ }\textbf
  {\bibinfo {volume} {32}},\ \bibinfo {pages} {2082} (\bibinfo {year}
  {1985})}\BibitemShut {NoStop}%
\bibitem [{\citenamefont {Eckhardt}(1995)}]{Eckhardt95}%
  \BibitemOpen
  \bibfield  {author} {\bibinfo {author} {\bibfnamefont {B.}~\bibnamefont
  {Eckhardt}},\ }\href@noop {} {\bibfield  {journal} {\bibinfo  {journal} {J.
  Phys. A}\ }\textbf {\bibinfo {volume} {28}},\ \bibinfo {pages} {3469}
  (\bibinfo {year} {1995})}\BibitemShut {NoStop}%
\bibitem [{\citenamefont {Hernandez}\ and\ \citenamefont
  {Miller}(1993)}]{Hernandez93}%
  \BibitemOpen
  \bibfield  {author} {\bibinfo {author} {\bibfnamefont {R.}~\bibnamefont
  {Hernandez}}\ and\ \bibinfo {author} {\bibfnamefont {W.~H.}\ \bibnamefont
  {Miller}},\ }\href {\doibase 10.1016/0009-2614(93)90071-8} {\bibfield
  {journal} {\bibinfo  {journal} {Chem. Phys. Lett.}\ }\textbf {\bibinfo
  {volume} {214}},\ \bibinfo {pages} {129} (\bibinfo {year}
  {1993})}\BibitemShut {NoStop}%
\bibitem [{\citenamefont {Hernandez}(1994)}]{Hernandez94}%
  \BibitemOpen
  \bibfield  {author} {\bibinfo {author} {\bibfnamefont {R.}~\bibnamefont
  {Hernandez}},\ }\href {\doibase 10.1063/1.467985} {\bibfield  {journal}
  {\bibinfo  {journal} {J. Chem. Phys.}\ }\textbf {\bibinfo {volume} {101}},\
  \bibinfo {pages} {9534} (\bibinfo {year} {1994})}\BibitemShut {NoStop}%
\bibitem [{\citenamefont {Jaff{\'e}}, \citenamefont {Farrelly},\ and\
  \citenamefont {Uzer}(1999)}]{Jaffe99}%
  \BibitemOpen
  \bibfield  {author} {\bibinfo {author} {\bibfnamefont {C.}~\bibnamefont
  {Jaff{\'e}}}, \bibinfo {author} {\bibfnamefont {D.}~\bibnamefont {Farrelly}},
  \ and\ \bibinfo {author} {\bibfnamefont {T.}~\bibnamefont {Uzer}},\
  }\href@noop {} {\bibfield  {journal} {\bibinfo  {journal} {Phys. Rev. A}\
  }\textbf {\bibinfo {volume} {60}},\ \bibinfo {pages} {3833} (\bibinfo {year}
  {1999})}\BibitemShut {NoStop}%
\bibitem [{\citenamefont {Jaff{\'e}}, \citenamefont {Farrelly},\ and\
  \citenamefont {Uzer}(2000)}]{Jaffe00}%
  \BibitemOpen
  \bibfield  {author} {\bibinfo {author} {\bibfnamefont {C.}~\bibnamefont
  {Jaff{\'e}}}, \bibinfo {author} {\bibfnamefont {D.}~\bibnamefont {Farrelly}},
  \ and\ \bibinfo {author} {\bibfnamefont {T.}~\bibnamefont {Uzer}},\
  }\href@noop {} {\bibfield  {journal} {\bibinfo  {journal} {Phys. Rev. Lett.}\
  }\textbf {\bibinfo {volume} {84}},\ \bibinfo {pages} {610} (\bibinfo {year}
  {2000})}\BibitemShut {NoStop}%
\bibitem [{\citenamefont {Koon}\ \emph {et~al.}(2000)\citenamefont {Koon},
  \citenamefont {Lo}, \citenamefont {Marsden},\ and\ \citenamefont
  {Ross}}]{Koon00}%
  \BibitemOpen
  \bibfield  {author} {\bibinfo {author} {\bibfnamefont {W.~S.}\ \bibnamefont
  {Koon}}, \bibinfo {author} {\bibfnamefont {M.~W.}\ \bibnamefont {Lo}},
  \bibinfo {author} {\bibfnamefont {J.~E.}\ \bibnamefont {Marsden}}, \ and\
  \bibinfo {author} {\bibfnamefont {S.~D.}\ \bibnamefont {Ross}},\ }\href
  {\doibase 10.1063/1.166509} {\bibfield  {journal} {\bibinfo  {journal}
  {Chaos}\ }\textbf {\bibinfo {volume} {10}},\ \bibinfo {pages} {427} (\bibinfo
  {year} {2000})}\BibitemShut {NoStop}%
\bibitem [{\citenamefont {Jaff{\'e}}\ \emph {et~al.}(2002)\citenamefont
  {Jaff{\'e}}, \citenamefont {Ross}, \citenamefont {Lo}, \citenamefont
  {Marsden}, \citenamefont {Farrelly},\ and\ \citenamefont {Uzer}}]{Jaffe02}%
  \BibitemOpen
  \bibfield  {author} {\bibinfo {author} {\bibfnamefont {C.}~\bibnamefont
  {Jaff{\'e}}}, \bibinfo {author} {\bibfnamefont {S.~D.}\ \bibnamefont {Ross}},
  \bibinfo {author} {\bibfnamefont {M.~W.}\ \bibnamefont {Lo}}, \bibinfo
  {author} {\bibfnamefont {J.}~\bibnamefont {Marsden}}, \bibinfo {author}
  {\bibfnamefont {D.}~\bibnamefont {Farrelly}}, \ and\ \bibinfo {author}
  {\bibfnamefont {T.}~\bibnamefont {Uzer}},\ }\href {\doibase
  10.1103/PhysRevLett.89.011101} {\bibfield  {journal} {\bibinfo  {journal}
  {Phys. Rev. Lett.}\ }\textbf {\bibinfo {volume} {89}},\ \bibinfo {pages}
  {011101} (\bibinfo {year} {2002})}\BibitemShut {NoStop}%
\bibitem [{\citenamefont {Uzer}\ \emph {et~al.}(2002)\citenamefont {Uzer},
  \citenamefont {Jaff{\'e}}, \citenamefont {Palaci{\'a}n}, \citenamefont
  {Yanguas},\ and\ \citenamefont {Wiggins}}]{Uzer02}%
  \BibitemOpen
  \bibfield  {author} {\bibinfo {author} {\bibfnamefont {T.}~\bibnamefont
  {Uzer}}, \bibinfo {author} {\bibfnamefont {C.}~\bibnamefont {Jaff{\'e}}},
  \bibinfo {author} {\bibfnamefont {J.}~\bibnamefont {Palaci{\'a}n}}, \bibinfo
  {author} {\bibfnamefont {P.}~\bibnamefont {Yanguas}}, \ and\ \bibinfo
  {author} {\bibfnamefont {S.}~\bibnamefont {Wiggins}},\ }\href {\doibase
  10.1088/0951-7715/15/4/301} {\bibfield  {journal} {\bibinfo  {journal}
  {Nonlinearity}\ }\textbf {\bibinfo {volume} {15}},\ \bibinfo {pages} {957}
  (\bibinfo {year} {2002})}\BibitemShut {NoStop}%
\bibitem [{\citenamefont {Komatsuzaki}\ and\ \citenamefont
  {Berry}(2002)}]{Komatsuzaki02}%
  \BibitemOpen
  \bibfield  {author} {\bibinfo {author} {\bibfnamefont {T.}~\bibnamefont
  {Komatsuzaki}}\ and\ \bibinfo {author} {\bibfnamefont {R.~S.}\ \bibnamefont
  {Berry}},\ }\href@noop {} {\bibfield  {journal} {\bibinfo  {journal} {Adv.
  Chem. Phys.}\ }\textbf {\bibinfo {volume} {123}},\ \bibinfo {pages} {79}
  (\bibinfo {year} {2002})}\BibitemShut {NoStop}%
\bibitem [{\citenamefont {Waalkens}, \citenamefont {Burbanks},\ and\
  \citenamefont {Wiggins}(2004{\natexlab{a}})}]{Waalkens04a}%
  \BibitemOpen
  \bibfield  {author} {\bibinfo {author} {\bibfnamefont {H.}~\bibnamefont
  {Waalkens}}, \bibinfo {author} {\bibfnamefont {A.}~\bibnamefont {Burbanks}},
  \ and\ \bibinfo {author} {\bibfnamefont {S.}~\bibnamefont {Wiggins}},\ }\href
  {\doibase 10.1088/0305-4470/37/24/L04} {\bibfield  {journal} {\bibinfo
  {journal} {J. Phys. A}\ }\textbf {\bibinfo {volume} {37}},\ \bibinfo {pages}
  {L257} (\bibinfo {year} {2004}{\natexlab{a}})}\BibitemShut {NoStop}%
\bibitem [{\citenamefont {Waalkens}, \citenamefont {Burbanks},\ and\
  \citenamefont {Wiggins}(2004{\natexlab{b}})}]{Waalkens04c}%
  \BibitemOpen
  \bibfield  {author} {\bibinfo {author} {\bibfnamefont {H.}~\bibnamefont
  {Waalkens}}, \bibinfo {author} {\bibfnamefont {A.}~\bibnamefont {Burbanks}},
  \ and\ \bibinfo {author} {\bibfnamefont {S.}~\bibnamefont {Wiggins}},\ }\href
  {\doibase 10.1063/1.1789891} {\bibfield  {journal} {\bibinfo  {journal} {J.
  Chem. Phys.}\ }\textbf {\bibinfo {volume} {121}},\ \bibinfo {pages} {6207}
  (\bibinfo {year} {2004}{\natexlab{b}})}\BibitemShut {NoStop}%
\bibitem [{\citenamefont {Pechukas}\ and\ \citenamefont
  {Pollak}(1979)}]{Pechukas79}%
  \BibitemOpen
  \bibfield  {author} {\bibinfo {author} {\bibfnamefont {P.}~\bibnamefont
  {Pechukas}}\ and\ \bibinfo {author} {\bibfnamefont {E.}~\bibnamefont
  {Pollak}},\ }\href {\doibase 10.1063/1.438575} {\bibfield  {journal}
  {\bibinfo  {journal} {J. Chem. Phys.}\ }\textbf {\bibinfo {volume} {71}},\
  \bibinfo {pages} {2062} (\bibinfo {year} {1979})}\BibitemShut {NoStop}%
\bibitem [{\citenamefont {Pollak}, \citenamefont {Child},\ and\ \citenamefont
  {Pechukas}(1980)}]{Pollak80}%
  \BibitemOpen
  \bibfield  {author} {\bibinfo {author} {\bibfnamefont {E.}~\bibnamefont
  {Pollak}}, \bibinfo {author} {\bibfnamefont {M.~S.}\ \bibnamefont {Child}}, \
  and\ \bibinfo {author} {\bibfnamefont {P.}~\bibnamefont {Pechukas}},\ }\href
  {\doibase 10.1063/1.439276} {\bibfield  {journal} {\bibinfo  {journal} {J.
  Chem. Phys.}\ }\textbf {\bibinfo {volume} {72}},\ \bibinfo {pages} {1669}
  (\bibinfo {year} {1980})}\BibitemShut {NoStop}%
\bibitem [{\citenamefont {Allahem}\ and\ \citenamefont
  {Bartsch}(2012)}]{Allahem12}%
  \BibitemOpen
  \bibfield  {author} {\bibinfo {author} {\bibfnamefont {A.}~\bibnamefont
  {Allahem}}\ and\ \bibinfo {author} {\bibfnamefont {T.}~\bibnamefont
  {Bartsch}},\ }\href {\doibase 10.1063/1.4769197} {\bibfield  {journal}
  {\bibinfo  {journal} {J. Chem. Phys.}\ }\textbf {\bibinfo {volume} {137}},\
  \bibinfo {pages} {214310} (\bibinfo {year} {2012})}\BibitemShut {NoStop}%
\bibitem [{\citenamefont {Garrett}\ and\ \citenamefont
  {Truhlar}(2005)}]{Garrett05a}%
  \BibitemOpen
  \bibfield  {author} {\bibinfo {author} {\bibfnamefont {B.~C.}\ \bibnamefont
  {Garrett}}\ and\ \bibinfo {author} {\bibfnamefont {D.~G.}\ \bibnamefont
  {Truhlar}},\ }in\ \href@noop {} {\emph {\bibinfo {booktitle} {Theory and
  Applications of Computational Chemistry: The First Forty Years}}},\ \bibinfo
  {editor} {edited by\ \bibinfo {editor} {\bibfnamefont {C.~E.}\ \bibnamefont
  {Dykstra}}, \bibinfo {editor} {\bibfnamefont {G.}~\bibnamefont {Frenking}},
  \bibinfo {editor} {\bibfnamefont {K.~S.}\ \bibnamefont {Kim}}, \ and\
  \bibinfo {editor} {\bibfnamefont {G.~E.}\ \bibnamefont {Scuseria}}}\
  (\bibinfo  {publisher} {Elsevier},\ \bibinfo {year} {2005})\ Chap.~\bibinfo
  {chapter} {5}, pp.\ \bibinfo {pages} {67--87}\BibitemShut {NoStop}%
\bibitem [{\citenamefont {Revuelta}\ \emph {et~al.}(2012)\citenamefont
  {Revuelta}, \citenamefont {Bartsch}, \citenamefont {Benito},\ and\
  \citenamefont {Borondo}}]{Revuelta12}%
  \BibitemOpen
  \bibfield  {author} {\bibinfo {author} {\bibfnamefont {F.}~\bibnamefont
  {Revuelta}}, \bibinfo {author} {\bibfnamefont {T.}~\bibnamefont {Bartsch}},
  \bibinfo {author} {\bibfnamefont {R.~M.}\ \bibnamefont {Benito}}, \ and\
  \bibinfo {author} {\bibfnamefont {F.}~\bibnamefont {Borondo}},\ }\href
  {\doibase 10.1063/1.3692182} {\bibfield  {journal} {\bibinfo  {journal} {J.
  Chem. Phys.}\ }\textbf {\bibinfo {volume} {136}},\ \bibinfo {pages} {091102}
  (\bibinfo {year} {2012})}\BibitemShut {NoStop}%
\bibitem [{\citenamefont {Bartsch}\ \emph {et~al.}(2012)\citenamefont
  {Bartsch}, \citenamefont {Revuelta}, \citenamefont {Benito},\ and\
  \citenamefont {Borondo}}]{Bartsch12}%
  \BibitemOpen
  \bibfield  {author} {\bibinfo {author} {\bibfnamefont {T.}~\bibnamefont
  {Bartsch}}, \bibinfo {author} {\bibfnamefont {F.}~\bibnamefont {Revuelta}},
  \bibinfo {author} {\bibfnamefont {R.~M.}\ \bibnamefont {Benito}}, \ and\
  \bibinfo {author} {\bibfnamefont {F.}~\bibnamefont {Borondo}},\ }\href
  {\doibase 10.1063/1.4726125} {\bibfield  {journal} {\bibinfo  {journal} {J.
  Chem. Phys.}\ }\textbf {\bibinfo {volume} {136}},\ \bibinfo {pages} {224510}
  (\bibinfo {year} {2012})}\BibitemShut {NoStop}%
\bibitem [{\citenamefont {Revuelta}\ \emph {et~al.}(2016)\citenamefont
  {Revuelta}, \citenamefont {Bartsch}, \citenamefont {Garcia-Muller},
  \citenamefont {Hernandez}, \citenamefont {Benito},\ and\ \citenamefont
  {Borondo}}]{Revuelta16}%
  \BibitemOpen
  \bibfield  {author} {\bibinfo {author} {\bibfnamefont {F.}~\bibnamefont
  {Revuelta}}, \bibinfo {author} {\bibfnamefont {T.}~\bibnamefont {Bartsch}},
  \bibinfo {author} {\bibfnamefont {P.~L.}\ \bibnamefont {Garcia-Muller}},
  \bibinfo {author} {\bibfnamefont {R.}~\bibnamefont {Hernandez}}, \bibinfo
  {author} {\bibfnamefont {R.~M.}\ \bibnamefont {Benito}}, \ and\ \bibinfo
  {author} {\bibfnamefont {F.}~\bibnamefont {Borondo}},\ }\href {\doibase
  10.1103/PhysRevE.93.062304} {\bibfield  {journal} {\bibinfo  {journal} {Phys.
  Rev. E}\ }\textbf {\bibinfo {volume} {93}},\ \bibinfo {pages} {062304}
  (\bibinfo {year} {2016})}\BibitemShut {NoStop}%
\bibitem [{\citenamefont {H{\"a}nggi}, \citenamefont {Talkner},\ and\
  \citenamefont {Borkovec}(1990)}]{Haenggi90}%
  \BibitemOpen
  \bibfield  {author} {\bibinfo {author} {\bibfnamefont {P.}~\bibnamefont
  {H{\"a}nggi}}, \bibinfo {author} {\bibfnamefont {P.}~\bibnamefont {Talkner}},
  \ and\ \bibinfo {author} {\bibfnamefont {M.}~\bibnamefont {Borkovec}},\
  }\href {\doibase 10.1103/RevModPhys.62.251} {\bibfield  {journal} {\bibinfo
  {journal} {Rev. Mod. Phys.}\ }\textbf {\bibinfo {volume} {62}},\ \bibinfo
  {pages} {251} (\bibinfo {year} {1990})}\BibitemShut {NoStop}%
\bibitem [{\citenamefont {Pechukas}(1976)}]{Pechukas76}%
  \BibitemOpen
  \bibfield  {author} {\bibinfo {author} {\bibfnamefont {P.}~\bibnamefont
  {Pechukas}},\ }in\ \href@noop {} {\emph {\bibinfo {booktitle} {Dynamics of
  Molecular Collisions, Part B}}},\ \bibinfo {editor} {edited by\ \bibinfo
  {editor} {\bibfnamefont {W.~H.}\ \bibnamefont {Miller}}}\ (\bibinfo
  {publisher} {Plenum},\ \bibinfo {address} {New York},\ \bibinfo {year}
  {1976})\ pp.\ \bibinfo {pages} {269--322}\BibitemShut {NoStop}%
\bibitem [{\citenamefont {Chandler}(1978)}]{Chandler78}%
  \BibitemOpen
  \bibfield  {author} {\bibinfo {author} {\bibfnamefont {D.}~\bibnamefont
  {Chandler}},\ }\href {\doibase 10.1063/1.436049} {\bibfield  {journal}
  {\bibinfo  {journal} {J. Chem. Phys.}\ }\textbf {\bibinfo {volume} {68}},\
  \bibinfo {pages} {2959} (\bibinfo {year} {1978})}\BibitemShut {NoStop}%
\bibitem [{\citenamefont {Bothma}, \citenamefont {Gilmore},\ and\ \citenamefont
  {McKenzie}(2010)}]{Bothma10}%
  \BibitemOpen
  \bibfield  {author} {\bibinfo {author} {\bibfnamefont {J.~P.}\ \bibnamefont
  {Bothma}}, \bibinfo {author} {\bibfnamefont {J.~B.}\ \bibnamefont {Gilmore}},
  \ and\ \bibinfo {author} {\bibfnamefont {R.~H.}\ \bibnamefont {McKenzie}},\
  }\href {\doibase 10.1088/1367-2630/12/5/055002} {\bibfield  {journal}
  {\bibinfo  {journal} {New J. Phys.}\ }\textbf {\bibinfo {volume} {12}},\
  \bibinfo {pages} {055002} (\bibinfo {year} {2010})}\BibitemShut {NoStop}%
\bibitem [{\citenamefont {Polli}\ \emph {et~al.}(2010)\citenamefont {Polli},
  \citenamefont {Alto{\`e}}, \citenamefont {Weingart}, \citenamefont
  {Spillane}, \citenamefont {Manzoni}, \citenamefont {Brida}, \citenamefont
  {Tomasello}, \citenamefont {Orlandi}, \citenamefont {Kukura}, \citenamefont
  {Mathies}, \citenamefont {Garavelli},\ and\ \citenamefont
  {Cerullo}}]{Polli10}%
  \BibitemOpen
  \bibfield  {author} {\bibinfo {author} {\bibfnamefont {D.}~\bibnamefont
  {Polli}}, \bibinfo {author} {\bibfnamefont {P.}~\bibnamefont {Alto{\`e}}},
  \bibinfo {author} {\bibfnamefont {O.}~\bibnamefont {Weingart}}, \bibinfo
  {author} {\bibfnamefont {K.~M.}\ \bibnamefont {Spillane}}, \bibinfo {author}
  {\bibfnamefont {C.}~\bibnamefont {Manzoni}}, \bibinfo {author} {\bibfnamefont
  {D.}~\bibnamefont {Brida}}, \bibinfo {author} {\bibfnamefont
  {G.}~\bibnamefont {Tomasello}}, \bibinfo {author} {\bibfnamefont
  {G.}~\bibnamefont {Orlandi}}, \bibinfo {author} {\bibfnamefont
  {P.}~\bibnamefont {Kukura}}, \bibinfo {author} {\bibfnamefont {R.~A.}\
  \bibnamefont {Mathies}}, \bibinfo {author} {\bibfnamefont {M.}~\bibnamefont
  {Garavelli}}, \ and\ \bibinfo {author} {\bibfnamefont {G.}~\bibnamefont
  {Cerullo}},\ }\href {\doibase 10.1038/nature09346} {\bibfield  {journal}
  {\bibinfo  {journal} {Nature}\ }\textbf {\bibinfo {volume} {467}},\ \bibinfo
  {pages} {440} (\bibinfo {year} {2010})}\BibitemShut {NoStop}%
\bibitem [{\citenamefont {Zwanzig}(1973)}]{Zwanzig73}%
  \BibitemOpen
  \bibfield  {author} {\bibinfo {author} {\bibfnamefont {R.}~\bibnamefont
  {Zwanzig}},\ }\href@noop {} {\bibfield  {journal} {\bibinfo  {journal} {J.
  Stat. Phys.}\ }\textbf {\bibinfo {volume} {9}},\ \bibinfo {pages} {215}
  (\bibinfo {year} {1973})}\BibitemShut {NoStop}%
\bibitem [{\citenamefont {Kramers}(1940)}]{Kramers40}%
  \BibitemOpen
  \bibfield  {author} {\bibinfo {author} {\bibfnamefont {H.~A.}\ \bibnamefont
  {Kramers}},\ }\href {\doibase 10.1016/S0031-8914(40)90098-2} {\bibfield
  {journal} {\bibinfo  {journal} {Physica (Utrecht)}\ }\textbf {\bibinfo
  {volume} {7}},\ \bibinfo {pages} {284} (\bibinfo {year} {1940})}\BibitemShut
  {NoStop}%
\bibitem [{\citenamefont {Grote}\ and\ \citenamefont {Hynes}(1980)}]{Grote80}%
  \BibitemOpen
  \bibfield  {author} {\bibinfo {author} {\bibfnamefont {R.~F.}\ \bibnamefont
  {Grote}}\ and\ \bibinfo {author} {\bibfnamefont {J.~T.}\ \bibnamefont
  {Hynes}},\ }\href@noop {} {\bibfield  {journal} {\bibinfo  {journal} {J.
  Chem. Phys.}\ }\textbf {\bibinfo {volume} {73}},\ \bibinfo {pages} {2715}
  (\bibinfo {year} {1980})}\BibitemShut {NoStop}%
\bibitem [{\citenamefont {Pollak}(1986)}]{Pollak86}%
  \BibitemOpen
  \bibfield  {author} {\bibinfo {author} {\bibfnamefont {E.}~\bibnamefont
  {Pollak}},\ }\href {\doibase 10.1063/1.451294} {\bibfield  {journal}
  {\bibinfo  {journal} {J. Chem. Phys.}\ }\textbf {\bibinfo {volume} {85}},\
  \bibinfo {pages} {865} (\bibinfo {year} {1986})}\BibitemShut {NoStop}%
\bibitem [{\citenamefont {Pollak}\ and\ \citenamefont
  {Talkner}(1993)}]{Pollak93a}%
  \BibitemOpen
  \bibfield  {author} {\bibinfo {author} {\bibfnamefont {E.}~\bibnamefont
  {Pollak}}\ and\ \bibinfo {author} {\bibfnamefont {P.}~\bibnamefont
  {Talkner}},\ }\href {\doibase 10.1103/PhysRevE.47.922} {\bibfield  {journal}
  {\bibinfo  {journal} {Phys. Rev. E}\ }\textbf {\bibinfo {volume} {47}},\
  \bibinfo {pages} {922} (\bibinfo {year} {1993})}\BibitemShut {NoStop}%
\bibitem [{\citenamefont {Talkner}\ and\ \citenamefont
  {Pollak}(1993)}]{Talkner93}%
  \BibitemOpen
  \bibfield  {author} {\bibinfo {author} {\bibfnamefont {P.}~\bibnamefont
  {Talkner}}\ and\ \bibinfo {author} {\bibfnamefont {E.}~\bibnamefont
  {Pollak}},\ }\href {\doibase 10.1103/PhysRevE.47.R21} {\bibfield  {journal}
  {\bibinfo  {journal} {Phys. Rev. E}\ }\textbf {\bibinfo {volume} {47}},\
  \bibinfo {pages} {R21} (\bibinfo {year} {1993})}\BibitemShut {NoStop}%
\bibitem [{\citenamefont {Talkner}(1994)}]{Talkner94a}%
  \BibitemOpen
  \bibfield  {author} {\bibinfo {author} {\bibfnamefont {P.}~\bibnamefont
  {Talkner}},\ }\href {\doibase 10.1016/0301-0104(93)E0426-V} {\bibfield
  {journal} {\bibinfo  {journal} {Chem. Phys.}\ }\textbf {\bibinfo {volume}
  {180}},\ \bibinfo {pages} {199} (\bibinfo {year} {1994})}\BibitemShut
  {NoStop}%
\bibitem [{\citenamefont {Bartsch}, \citenamefont {Hernandez},\ and\
  \citenamefont {Uzer}(2005)}]{Bartsch05b}%
  \BibitemOpen
  \bibfield  {author} {\bibinfo {author} {\bibfnamefont {T.}~\bibnamefont
  {Bartsch}}, \bibinfo {author} {\bibfnamefont {R.}~\bibnamefont {Hernandez}},
  \ and\ \bibinfo {author} {\bibfnamefont {T.}~\bibnamefont {Uzer}},\ }\href
  {\doibase 10.1103/PhysRevLett.95.058301} {\bibfield  {journal} {\bibinfo
  {journal} {Phys. Rev. Lett.}\ }\textbf {\bibinfo {volume} {95}},\ \bibinfo
  {pages} {058301} (\bibinfo {year} {2005})}\BibitemShut {NoStop}%
\bibitem [{\citenamefont {Bartsch}, \citenamefont {Uzer},\ and\ \citenamefont
  {Hernandez}(2005)}]{Bartsch05c}%
  \BibitemOpen
  \bibfield  {author} {\bibinfo {author} {\bibfnamefont {T.}~\bibnamefont
  {Bartsch}}, \bibinfo {author} {\bibfnamefont {T.}~\bibnamefont {Uzer}}, \
  and\ \bibinfo {author} {\bibfnamefont {R.}~\bibnamefont {Hernandez}},\ }\href
  {\doibase 10.1063/1.2109827} {\bibfield  {journal} {\bibinfo  {journal} {J.
  Chem. Phys.}\ }\textbf {\bibinfo {volume} {123}},\ \bibinfo {pages} {204102}
  (\bibinfo {year} {2005})}\BibitemShut {NoStop}%
\bibitem [{\citenamefont {Bartsch}\ \emph {et~al.}(2006)\citenamefont
  {Bartsch}, \citenamefont {Uzer}, \citenamefont {Moix},\ and\ \citenamefont
  {Hernandez}}]{Bartsch06a}%
  \BibitemOpen
  \bibfield  {author} {\bibinfo {author} {\bibfnamefont {T.}~\bibnamefont
  {Bartsch}}, \bibinfo {author} {\bibfnamefont {T.}~\bibnamefont {Uzer}},
  \bibinfo {author} {\bibfnamefont {J.~M.}\ \bibnamefont {Moix}}, \ and\
  \bibinfo {author} {\bibfnamefont {R.}~\bibnamefont {Hernandez}},\ }\href
  {\doibase 10.1063/1.2206587} {\bibfield  {journal} {\bibinfo  {journal} {J.
  Chem. Phys.}\ }\textbf {\bibinfo {volume} {124}},\ \bibinfo {pages} {244310}
  (\bibinfo {year} {2006})}\BibitemShut {NoStop}%
\bibitem [{\citenamefont {Bartsch}\ \emph {et~al.}(2008)\citenamefont
  {Bartsch}, \citenamefont {Uzer}, \citenamefont {Moix},\ and\ \citenamefont
  {Hernandez}}]{Bartsch08}%
  \BibitemOpen
  \bibfield  {author} {\bibinfo {author} {\bibfnamefont {T.}~\bibnamefont
  {Bartsch}}, \bibinfo {author} {\bibfnamefont {T.}~\bibnamefont {Uzer}},
  \bibinfo {author} {\bibfnamefont {J.~M.}\ \bibnamefont {Moix}}, \ and\
  \bibinfo {author} {\bibfnamefont {R.}~\bibnamefont {Hernandez}},\ }\href
  {\doibase 10.1021/jp0755600} {\bibfield  {journal} {\bibinfo  {journal} {J.
  Phys. Chem. B}\ }\textbf {\bibinfo {volume} {112}},\ \bibinfo {pages} {206}
  (\bibinfo {year} {2008})}\BibitemShut {NoStop}%
\bibitem [{\citenamefont {Hernandez}, \citenamefont {Uzer},\ and\ \citenamefont
  {Bartsch}(2010)}]{Hernandez10}%
  \BibitemOpen
  \bibfield  {author} {\bibinfo {author} {\bibfnamefont {R.}~\bibnamefont
  {Hernandez}}, \bibinfo {author} {\bibfnamefont {T.}~\bibnamefont {Uzer}}, \
  and\ \bibinfo {author} {\bibfnamefont {T.}~\bibnamefont {Bartsch}},\ }\href
  {\doibase 10.1016/j.chemphys.2010.01.016} {\bibfield  {journal} {\bibinfo
  {journal} {Chemical Physics}\ }\textbf {\bibinfo {volume} {370}},\ \bibinfo
  {pages} {270 } (\bibinfo {year} {2010})}\BibitemShut {NoStop}%
\bibitem [{\citenamefont {Hershkovitz}(1998)}]{Hershkovitz98}%
  \BibitemOpen
  \bibfield  {author} {\bibinfo {author} {\bibfnamefont {E.}~\bibnamefont
  {Hershkovitz}},\ }\href {\doibase 10.1063/1.476380} {\bibfield  {journal}
  {\bibinfo  {journal} {J. Chem. Phys.}\ }\textbf {\bibinfo {volume} {108}},\
  \bibinfo {pages} {9253} (\bibinfo {year} {1998})}\BibitemShut {NoStop}%
\bibitem [{\citenamefont {Hershkovitz}\ and\ \citenamefont
  {Hernandez}(2001)}]{Hershkovitz01}%
  \BibitemOpen
  \bibfield  {author} {\bibinfo {author} {\bibfnamefont {E.}~\bibnamefont
  {Hershkovitz}}\ and\ \bibinfo {author} {\bibfnamefont {R.}~\bibnamefont
  {Hernandez}},\ }\href {\doibase 10.1021/jp0037044} {\bibfield  {journal}
  {\bibinfo  {journal} {J. Phys. Chem. A}\ }\textbf {\bibinfo {volume} {105}},\
  \bibinfo {pages} {2687} (\bibinfo {year} {2001})}\BibitemShut {NoStop}%
\bibitem [{\citenamefont {Garc{\'\i}a-M{\"u}ller}\ \emph
  {et~al.}(2008)\citenamefont {Garc{\'\i}a-M{\"u}ller}, \citenamefont
  {Borondo}, \citenamefont {Hernandez},\ and\ \citenamefont
  {Benito}}]{Muller10}%
  \BibitemOpen
  \bibfield  {author} {\bibinfo {author} {\bibfnamefont {P.~L.}\ \bibnamefont
  {Garc{\'\i}a-M{\"u}ller}}, \bibinfo {author} {\bibfnamefont {F.}~\bibnamefont
  {Borondo}}, \bibinfo {author} {\bibfnamefont {R.}~\bibnamefont {Hernandez}},
  \ and\ \bibinfo {author} {\bibfnamefont {R.~M.}\ \bibnamefont {Benito}},\
  }\href {\doibase 10.1103/PhysRevLett.101.178302} {\bibfield  {journal}
  {\bibinfo  {journal} {Phys. Rev. Lett.}\ }\textbf {\bibinfo {volume} {101}},\
  \bibinfo {pages} {178302} (\bibinfo {year} {2008})}\BibitemShut {NoStop}%
\bibitem [{\citenamefont {Ferrario}\ and\ \citenamefont
  {Grigolini}(1979)}]{Ferrario79}%
  \BibitemOpen
  \bibfield  {author} {\bibinfo {author} {\bibfnamefont {M.}~\bibnamefont
  {Ferrario}}\ and\ \bibinfo {author} {\bibfnamefont {P.}~\bibnamefont
  {Grigolini}},\ }\href {\doibase 10.1063/1.524019} {\bibfield  {journal}
  {\bibinfo  {journal} {J. Math. Phys.}\ }\textbf {\bibinfo {volume} {20}},\
  \bibinfo {pages} {2567} (\bibinfo {year} {1979})}\BibitemShut {NoStop}%
\bibitem [{\citenamefont {Grigolini}(1982)}]{Grigolini82}%
  \BibitemOpen
  \bibfield  {author} {\bibinfo {author} {\bibfnamefont {P.}~\bibnamefont
  {Grigolini}},\ }\href {\doibase 10.1007/BF01008940} {\bibfield  {journal}
  {\bibinfo  {journal} {Journal of Statistical Physics}\ }\textbf {\bibinfo
  {volume} {27}},\ \bibinfo {pages} {283} (\bibinfo {year} {1982})}\BibitemShut
  {NoStop}%
\bibitem [{\citenamefont {Marchesoni}\ and\ \citenamefont
  {Grigolini}(1983)}]{Marchesoni83}%
  \BibitemOpen
  \bibfield  {author} {\bibinfo {author} {\bibfnamefont {F.}~\bibnamefont
  {Marchesoni}}\ and\ \bibinfo {author} {\bibfnamefont {P.}~\bibnamefont
  {Grigolini}},\ }\href {\doibase 10.1063/1.444554} {\bibfield  {journal}
  {\bibinfo  {journal} {J. Chem. Phys.}\ }\textbf {\bibinfo {volume} {78}},\
  \bibinfo {pages} {6287} (\bibinfo {year} {1983})}\BibitemShut {NoStop}%
\bibitem [{\citenamefont {Martens}(2002)}]{Martens02}%
  \BibitemOpen
  \bibfield  {author} {\bibinfo {author} {\bibfnamefont {C.~C.}\ \bibnamefont
  {Martens}},\ }\href {\doibase 10.1063/1.1436116} {\bibfield  {journal}
  {\bibinfo  {journal} {J. Chem. Phys.}\ }\textbf {\bibinfo {volume} {116}},\
  \bibinfo {pages} {2516} (\bibinfo {year} {2002})}\BibitemShut {NoStop}%
\bibitem [{\citenamefont {Bartsch}(2009)}]{Bartsch09}%
  \BibitemOpen
  \bibfield  {author} {\bibinfo {author} {\bibfnamefont {T.}~\bibnamefont
  {Bartsch}},\ }\href {\doibase 10.1063/1.3239473} {\bibfield  {journal}
  {\bibinfo  {journal} {J. Chem. Phys.}\ }\textbf {\bibinfo {volume} {131}},\
  \bibinfo {pages} {124121} (\bibinfo {year} {2009})}\BibitemShut {NoStop}%
\bibitem [{\citenamefont {Kawai}\ \emph {et~al.}(2007)\citenamefont {Kawai},
  \citenamefont {Bandrauk}, \citenamefont {Jaff{\'e}}, \citenamefont {Bartsch},
  \citenamefont {Palaci{\'a}n},\ and\ \citenamefont {Uzer}}]{Kawai07a}%
  \BibitemOpen
  \bibfield  {author} {\bibinfo {author} {\bibfnamefont {S.}~\bibnamefont
  {Kawai}}, \bibinfo {author} {\bibfnamefont {A.~D.}\ \bibnamefont {Bandrauk}},
  \bibinfo {author} {\bibfnamefont {C.}~\bibnamefont {Jaff{\'e}}}, \bibinfo
  {author} {\bibfnamefont {T.}~\bibnamefont {Bartsch}}, \bibinfo {author}
  {\bibfnamefont {J.}~\bibnamefont {Palaci{\'a}n}}, \ and\ \bibinfo {author}
  {\bibfnamefont {T.}~\bibnamefont {Uzer}},\ }\href {\doibase
  10.1063/1.2720841} {\bibfield  {journal} {\bibinfo  {journal} {J. Chem.
  Phys.}\ }\textbf {\bibinfo {volume} {126}},\ \bibinfo {pages} {164306}
  (\bibinfo {year} {2007})}\BibitemShut {NoStop}%
\bibitem [{\citenamefont {Isserlis}(1916)}]{Isserlis16}%
  \BibitemOpen
  \bibfield  {author} {\bibinfo {author} {\bibfnamefont {L.}~\bibnamefont
  {Isserlis}},\ }\href@noop {} {\bibfield  {journal} {\bibinfo  {journal}
  {Biometrika}\ }\textbf {\bibinfo {volume} {11}},\ \bibinfo {pages} {185}
  (\bibinfo {year} {1916})}\BibitemShut {NoStop}%
\bibitem [{\citenamefont {Isserlis}(1918)}]{Isserlis18}%
  \BibitemOpen
  \bibfield  {author} {\bibinfo {author} {\bibfnamefont {L.}~\bibnamefont
  {Isserlis}},\ }\href@noop {} {\bibfield  {journal} {\bibinfo  {journal}
  {Biometrika}\ }\textbf {\bibinfo {volume} {12}},\ \bibinfo {pages} {134}
  (\bibinfo {year} {1918})}\BibitemShut {NoStop}%
\bibitem [{\citenamefont {Wolfram}(2005)}]{Mathematica}%
  \BibitemOpen
  \bibfield  {author} {\bibinfo {author} {\bibfnamefont {S.}~\bibnamefont
  {Wolfram}},\ }\href@noop {} {\emph {\bibinfo {title} {The Mathematica
  Book}}}\ (\bibinfo  {publisher} {Wolfram Media},\ \bibinfo {year}
  {2005})\BibitemShut {NoStop}%
\bibitem [{\citenamefont {Tipler}\ and\ \citenamefont
  {Mosca}(2007)}]{Tipler07}%
  \BibitemOpen
  \bibfield  {author} {\bibinfo {author} {\bibfnamefont {P.~A.}\ \bibnamefont
  {Tipler}}\ and\ \bibinfo {author} {\bibfnamefont {G.~P.}\ \bibnamefont
  {Mosca}},\ }\href@noop {} {\emph {\bibinfo {title} {Physics for Scientists
  and Engineers}}},\ \bibinfo {edition} {6th}\ ed.,\ Vol.~\bibinfo {volume}
  {1}\ (\bibinfo  {publisher} {W. H. Freeman},\ \bibinfo {address} {New York},\
  \bibinfo {year} {2007})\BibitemShut {NoStop}%
  \BibitemOpen
  \href@noop {} {}\bibinfo {note} {See explicit expressions of the transmission
  factor in the \textit{Mathematica} notebook available as Supplemental
  Material}\BibitemShut {NoStop}%
\end{thebibliography}

%

\end{document}